\pdfoutput=1
\documentclass[prd,preprint,eqsecnum,nofootinbib,amsmath,amssymb,preprintnumbers,tightenlines,longbibliography]{revtex4-1}

\usepackage{afterpage}
\usepackage{mathrsfs}
\usepackage{graphicx}% Include figure files
\usepackage{bm}% bold math
\usepackage{paralist}
\usepackage[colorlinks,linkcolor=blue,citecolor=blue]{hyperref}

\def\t{\tilde}

\def\e{{\epsilon} }

\def\P{{\mathcal P}}
\def\V{{\mathcal V}}
\def\K{{\mathcal K}}
\def\G{{\mathcal G}}

\def\kx{{k_\xi}}

\def\Eq#1{Eq.~(\ref{#1})}
\def\Eqs#1{Eqs.~(\ref{#1})}

\def\Fig#1{Fig.~\ref{#1}}
\def\Sect#1{Section~\ref{#1}}

\def\bra{\langle}
\def\ket{\rangle}

\def\sech{\mbox{sech}}

\def\be{\begin{equation}}
\def\ee{\end{equation}}
\def\bea{\begin{eqnarray}}
\def\eea{\end{eqnarray}}

\newcommand\bal{\begin{align}}
\newcommand\eal{\end{align}}

\begin{document}

\title{Hydrodynamical noise and Gubser flow}

\author{Li Yan} 
\email{li.yan@cea.fr}
\author{Hanna Gr\"onqvist}
\email{hanna.gronqvist@cea.fr}
\affiliation
    {%
%	CNRS, URA2306, IPhT, 
%	Institut de Physique Th\'eorique de Saclay, F-91191
%	Gif-sur-Yvette, France \\
	Institut de Physique Th{\'e}orique, Universit\'e Paris Saclay, 
        CEA, CNRS, 
	F-91191 Gif-sur-Yvette, France\\
    }%

\begin{abstract}

Hydrodynamical noise is introduced on top of Gubser's analytical solution
to viscous hydrodynamics. With respect to the ultra-central collision events
of Pb-Pb, p-Pb and p-p at the LHC energies, we solve the evolution of
noisy fluid systems and calculate 
the radial flow velocity correlations. We show 
that the absolute amplitude of the hydrodynamical noise is determined by the
multiplicity of the collision event. The 
evolution of azimuthal anisotropies, which is related to the 
generation of harmonic flow, receives finite enhancements from hydrodynamical
noise. Although it
is strongest in the p-p systems, 
the effect of hydrodynamical noise on flow harmonics is found to be
negligible, especially in the ultra-central Pb-Pb collisions.
For the short-range correlations, hydrodynamical noise contributes to the 
formation of a near-side peak on top of the correlation structure originated from
initial state fluctuations. The shape of the peak is affected by
the strength of hydrodynamical noise, whose height and width grow from the Pb-Pb
system to the p-Pb and p-p systems.

\end{abstract}

\maketitle

\section{introduction}

One of the recent focuses on relativistic heavy-ion collisions carried out at the
Large Hadron Collider (LHC) and Relativistic Heavy-Ion Collider (RHIC) % is collectivity (RHIC)
lies in small colliding systems, including p-p and p-Pb at the 
LHC~\cite{CMS:2015zpa,Aad:2015gqa,Chatrchyan:2013nka,Aad:2014lta,CMS:2012qk,Khachatryan:2015waa}, 
d-Au and $^3$He-Au at RHIC~\cite{greenspan1996,Adare:2013piz}. 
Especially in the collision events with sufficiently high multiplicity, it has been noticed 
that observables related to multi-particle correlations are consistent
with a medium collective expansion scenario~\cite{Bzdak:2013rya,Yan:2013laa,Khachatryan:2015waa}. 
%which is strongly supported by the measurements of collective flow.
%%and the so-called collective flow has been measured. 
%For instance, elliptic flow $v_2$~\cite{Ollitrault:1992bk} in p-Pb 
%was found very similar to that in Pb-Pb collision events of comparable 
%multiplicity~\cite{Basar:2013hea,Chatrchyan:2013nka}.
%Moreover, $v_2$ in p-Pb measured from multi-particle correlations presents a pattern 
%which quantitatively agrees with predictions 
%from initial ellipticity fluctuations~\cite{Bzdak:2013rya,Yan:2013laa,Khachatryan:2015waa}.  
%Despite alternative interpretations of collectivity
%from pre-equilibrium physics~\cite{Gyulassy:2014cfa}, 
Relativistic hydrodynamics is a natural candidate to simulate and investigate
the collective
expansion of a QCD medium, even in small 
colliding systems
\cite{Kozlov:2014fqa,Qin:2013bha,Bozek:2011if,Nagle:2013lja,Bozek:2013uha,Bozek:2014cya,Yan:2015fva}. 
However, applying viscous hydrodynamics to small colliding
systems is challenged by a couple of factors. First, 
the applicability of viscous hydrodynamics is constrained by
the convergence of gradient expansion, which is normally quantified by 
%dissipations in
%a medium is effectively characterized in viscous hydrodynamics via gradient
%expansion. Convergence of the gradient expansion, namely, the applicability
%of viscous hydrodynamics, % in a system, 
%is controlled by 
%the ratio between microscopic length scale and macroscopic length scale, 
%which is normally quantified by 
Knudsen number Kn.
%Normally, we
%define the ratio as Knudsen number $K_n\sim\lambda_{\mbox{\tiny mft}}/L$,
%with $\lambda_{\mbox{\tiny mft}}$ the mean free path and $L$ the system
%size. 
In smaller systems, the smaller system size leads to a larger value of Kn
and the applicability of hydrodynamics is deteriorated accordingly~\cite{Niemi:2014wta}.
Second, %effects due to 
although hydrodynamical noise was introduced long ago by
Landau and Lifshitz~\cite{Landau-sp1,Landau-sp2}, 
and was recently generalized to relativistic systems 
by Kapusta, M\"uller and Stephanov~\cite{Kapusta:2011gt}, 
it is neglected in 
most of the present hydrodynamic simulations. 
%have neglected thermal fluctuations. 
Since
hydrodynamical noise is generically 
associated with dissipations and expected to be
more pronounced in small systems, its influence in the small 
colliding systems needs to be clarified. The purpose of this work is to 
quantitatively estimate the effects of thermal noise in 
Pb-Pb, p-Pb and p-p collisions at the LHC energies, in order to
test the applicability of \emph{noisy} and \emph{viscous}
hydrodynamics in these systems.

Instead of more explicit numerical
simulations of hydrodynamics with thermal noise~\cite{Young:2014pka,MuraseQM15}, 
this work resorts to %the recent work of Kapusta, M\"uller and Stephanov~\cite{},
an analytical solution of viscous hydrodynamics given by Gubser and
Yarom~\cite{Gubser:2010ze,Gubser:2010ui}, known as the Gubser flow. 
%Effects of fluctuations in a system
%are meaningful on a statistical level, which demands
%a large mount of events in numerical simulations. Accounting for the fact
%that in heavy-ion collisions initial state (quantum) fluctuations
%fluctuations must be implemented in hydro modelings with sufficiently 
By doing so, the inclusion of thermal noise in hydrodynamics 
is simplified dramatically
%on top of an analytical solvable hydrodynamics 
both theoretically and numerically. % especially, accounting for the fact that
%initial state (quantum) fluctuations demand hydro simulations on an
%event-by-event basis. 
Although Gubser flow is not as realistic as being required 
for heavy-ion collisions, it mimics the expansion of a hot conformal medium after
its thermalization. Therefore, as one preliminary work on hydrodynamical noise, 
we will restrict 
ourselves to the initial stage of heavy-ion collisions. We will not address
hadronization and freeze-out.  

%Throughout this work, we take a $(-,+,+,+)$ matrix convention, with quantities 
%defined accordingly. For instance, the projection operator $\Delta^{\mu\nu}$
%is defined through flow velocity $u^\mu$ as
%\be
%\Delta^{\mu\nu}=u^\mu u^\nu + g^{\mu\nu}\,,
%\ee 
%which further defines spatial gradient $\nabla^\mu = \Delta^{\mu\nu}\partial_\nu$.
%Tensor indices within angular brackets are transverse, traceless and
%symmetric, which can be defined in terms of projection operator, for instance, 
%\be
%\nabla^{\bra\mu} u^{\nu\ket}=
%\frac{1}{2}\Delta^{\mu\alpha}\Delta^{\nu\beta}
%\left[\nabla_\alpha u_\beta+\nabla_\beta u_\alpha\right]
%-\frac{1}{3}\Delta^{\mu\nu}\Delta^{\alpha\beta} \nabla_\alpha u_\beta\,.
%\ee
%We also use angular brackets $\bra\ldots\ket$
%around a quantity to denote ensemble average to take care of the effects of
%thermal fluctuations. More precisely,
%in our simulations ensemble average is obtained from a number of events with
%the same multiplicity as, 
%\be
%\bra\ldots\ket=\frac{1}{N_{\mbox{\tiny event}}}\sum \ldots\,,
%\ee
%except when initial state fluctuations are taken into account
%as well in \Sect{sec:formal-solution}. 

This paper is organized as follows. The theoretical framework of viscous
hydrodynamics and hydrodynamical noise is discussed on a general ground 
in \Sect{sec:hydro_fluc}.
In \Sect{sec:gubser} we briefly review Gubser flow, then thermal noise is 
introduced into Gubser's solution in \Sect{sec:noisy-gubser} in parallel to the
case of 1+1D Bjorken flow~\cite{Kapusta:2011gt}. In order to estimate 
the effects of hydrodynamical noise with respect to the Pb-Pb, p-Pb and p-p systems, 
we solve the noisy Gubser flow in \Sect{sec:solution}, 
with discussions in terms of %based on 
formal solutions presented in \Sect{sec:formal-solution}
and numerical simulations in \Sect{sec:num-solution}.
Summary and conclusions are given in \Sect{sec:summary}.

\section{Hydrodynamics and hydrodynamical fluctuations}
\label{sec:hydro_fluc}

In this section and throughout this work, we work with a metric signature
that is mostly positive $(-,+,+,+)$,
thereby the flow velocity $u^\mu$ is normalized as $u^2=-1$, 
and the projection operator $\Delta^{\mu\nu}$ is defined as
$
\Delta^{\mu\nu}=u^\mu u^\nu + g^{\mu\nu}\,.
$ 
%which further defines spatial gradient $\nabla^\mu = \Delta^{\mu\nu}\partial_\nu$.
Tensor indices within angular brackets are transverse, traceless and
symmetric, %which can be defined in terms of projection operator, 
%for instance, 
%\be
%\nabla^{\bra\mu} u^{\nu\ket}=
%\frac{1}{2}\Delta^{\mu\alpha}\Delta^{\nu\beta}
%\left[\nabla_\alpha u_\beta+\nabla_\beta u_\alpha\right]
%-\frac{1}{3}\Delta^{\mu\nu}\Delta^{\alpha\beta} \nabla_\alpha u_\beta\,,
%\ee
while tensor indices inside parentheses are symmetric.
Except being specified as in \Sect{sec:formal-solution}, 
we also use angular brackets around a quantity 
to denote ensemble average, which is defined as the average over
simulation events with respect to the same initial condition.
% with respect to collision events of the same total multiplicity.

The evolution of quark-gluon plasma in heavy-ion collisions 
can be well described by relativistic hydrodynamics,
together with an equation of state (EOS) originated from lattice QCD. 
Neglecting baryon number conservation, hydrodynamics
regarding heavy-ion collisions
is formulated as the conservation of energy-momentum,
\be
\label{equ:hydro-eom}
d_\mu T^{\mu\nu}=0\,,
\ee 
where we have taken $d_\mu$ to indicate the
covariant derivative and $T^{\mu\nu}$ is the energy-momentum tensor. In the Landau frame,
\be
T^{\mu\nu} = \e u^\mu u^\nu + \P\Delta^{\mu\nu} + \Pi^{\mu\nu}\,,
\ee
with dissipative effects characterized by the stress tensor $\Pi^{\mu\nu}$. 
Viscous hydrodynamics generally takes 
a form of gradient expansion, 
\be
\label{equ:visc-eom1}
\Pi^{\mu\nu}=-\eta \sigma^{\mu\nu}%2\nabla^{\bra \mu} u^{\nu\ket}
-\zeta\Delta^{\mu\nu} \nabla\cdot u+O(\nabla^2)\,,
\ee
where $\sigma^{\mu\nu}=2\nabla^{\bra \mu} u^{\nu\ket}$ and $\nabla^\mu\equiv \Delta^{\mu\nu}d_\nu$. 
$\Pi^{\mu\nu}$ has a determined form
up to first order in the expansion, which is known as the Navier-Stokes hydrodynamics. 
Throughout this work we shall only consider conformal fluids with linear equation of state
$\e=3\P$, while bulk viscous effects are ignored
by taking bulk viscosity $\zeta=0$. 
%Higher order viscous corrections are taken into account order by order with respect to a gradient
%expansion. For instance, for conformal fluids, one has up to second order in the gradient expansion,
%\begin{align}
%\label{equ:visc-eom2}
%\Pi^{\mu\nu}=&-\eta\sigma^{\mu\nu}\nonumber\\
%&+\eta\tau_\Pi\left[\bra D\sigma^{\mu\nu}\ket + \frac{1}{3}\sigma^{\mu\nu}\nabla\cdot u\right]
%+\kappa\left[R^{\bra\mu\nu\ket}-2u_\alpha R^{\alpha\bra \mu\nu\ket\beta}u_\beta\right]\nonumber\\
%&+\lambda_1\sigma^{\bra\mu}_\lambda\sigma^{\nu\ket\lambda}	
%+\lambda_2\sigma^{\bra\mu}_\lambda\Omega^{\nu\ket\lambda}
%+\lambda_3\Omega^{\bra\mu}_\lambda\Omega^{\nu\ket\lambda}\,.
%\end{align}
When applying hydrodynamics to heavy-ion collision, $\Pi^{\mu\nu}$ in \Eq{equ:visc-eom1} can
be practically treated as one dynamical quantity, and accordingly \Eq{equ:visc-eom1} is recognized as
its equation of motion. 
%It is also worth mentioning that dissipations result in entropy production, 
%which can be expressed via $\Pi^{\mu\nu}$ as 
%\be
%\frac{dS}{dt}=-\int d^3 \vec x \frac{\nabla_{(\mu}u_{\nu)}}{T}\Pi^{\mu\nu}
%\ee 

A standard way of introducing thermal noise in hydrodynamics was suggested
by Landau and Lifshitz~\cite{Landau-sp1,Landau-sp2} for the 
first order dissipative hydrodynamics, and extended to
the framework of relativistic hydrodynamics by Kapusta, M\"uller and Stephanov~\cite{Kapusta:2011gt}. 
If one associates for each thermodynamical 
quantity with a fluctuation term which characterizes thermal noise, e.g., temperature,
energy density and pressure are expressed as
(suffix `0' indicates ensemble-averaged quantity.)
\begin{subequations}
\label{equ:thermal-noise}
\begin{align}
T(x)  =&T_0(x) + \delta T(x)\,,\\
\e(x )=&\e_0(x) + \delta \e(x)\,,\\
\P(x) =&\P_0(x) + \delta \P(x)\,,
\end{align}
\end{subequations}
and so for each dynamical quantity in the fluid system, 
\begin{subequations}
\label{equ:hydro-noise}
\begin{align}
u^\mu(x) = &u_0^\mu(x) + \delta u^\mu(x)\,,\\
\Pi^{\mu\nu}(x) = & \Pi_0^{\mu\nu}(x) + S^{\mu\nu}(x)\,.
\end{align}
\end{subequations}
Note that the hydrodynamical noise term $S^{\mu\nu}$ is introduced with respect to $\Pi^{\mu\nu}$.
One thus arrives at a new expression for the energy-momentum tensor,
\be
\label{equ:Tmn}
T^{\mu\nu}=T_0^{\mu\nu} + \delta T^{\mu\nu}\,.
\ee
$T_0^{\mu\nu}$ in \Eq{equ:Tmn} indicates contributions from ensemble-averaged hydro quantities, 
while $\delta T^{\mu\nu}$ is determined by thermal fluctuations. 
Apart from cases, such as those where %instability or 
phase transition plays an significant role in the system evolution
(cf.~\cite{Stephanov:1999zu}), which 
is beyond the scope of this work, thermal fluctuations are relatively small variables. 
Accordingly, one can treat thermal fluctuations perturbatively
with respect to the background evolution $d_\mu T^{\mu\nu}_0=0$. 
To the first order in fluctuations, 
hydro equations of motion for fluctuations $d_\mu \delta T^{\mu\nu}=0$
are linearized and lead to
\begin{subequations}
\label{equ:fluc-eom2}
\begin{align}
\delta w Du_\alpha + w\delta u^\mu d_\mu u_\alpha + (D w  + w\partial\cdot u) \delta u_\alpha 
+ \nabla_\alpha\delta \P + wD\delta u_\alpha + d_\mu(\delta \Pi_\alpha^\mu
+S_{\alpha}^\mu)=&0\\
D\delta \e + \delta w\partial\cdot u + d_\mu( w \delta u^\mu)
+w\delta u^\alpha D u_\alpha - u^\alpha d_\mu(\delta \Pi^\mu_\alpha + S^\mu_\alpha)=&0
\end{align}
\end{subequations}
where $w=\e+\P$ is the enthalpy density. %{\color{blue} about $\delta \Pi$. ..}
$\delta \Pi^{\mu\nu}$ in \Eqs{equ:fluc-eom2} is a term induced by $\delta T$ and $\delta u^\mu$.
Without the hydrodynamical noise term $S^{\mu\nu}$, \Eqs{equ:fluc-eom2} also describe
the hydro evolution of initial state fluctuations to linear order. 
In \Eqs{equ:fluc-eom2} and in the following, we ignore 
the suffix `0' for the ensemble-averaged quantities for simplicity, when confusions 
do not arise.

When the system is in thermal equilibrium, two-point autocorrelations of 
these fluctuations in \Eqs{equ:thermal-noise}
and $\delta u^\mu$ are known to be local in space and time, with correlation strength 
constrained by thermodynamical
variables in equilibrium~\cite{Landau-sp1,Landau-sp2}. 
%(Note that in this case $\Pi^{\mu\nu}$, as well as $S^{\mu\nu}$, precisely vanish.) 
%For instance, the equal-time two-point correlation of $\delta T$ is~\cite{} 
%\be
%\bra \delta T(t, \vec x)\delta T(t, \vec x')\ket=\frac{T^2}{n c_v}\delta^{(3)}(\vec x -\vec x')\,,
%\ee
%where $n$ is density and $c_v$ is specific heat per unit mass of the medium. 
Autocorrelations of thermal quantities
are related with each other through the 
equation of state. When the system is out-of-equilibrium and
driven by hydrodynamics, 
especially when dissipation is taken into account, the autocorrelation of these fluctuations  
must be determined with respect to that of hydrodynamical noise $S^{\mu\nu}$, 
according to the fluctuation-dissipation theorem, 
\be
\label{equ:fd-theorem}
\bra S^{\mu\nu}(x_1)S^{\alpha\beta}(x_2)\ket= (\gamma^{\mu\nu\alpha\beta}+\gamma^{\alpha\beta\mu\nu})
\delta^{(4)}(x_1-x_2)\,,
\ee
where the Onsager coefficient $\gamma^{\mu\nu\alpha\beta}$ relates  
the symmetric tensor $\nabla^{(\mu} u^{\nu)}/T$ to %the traceless, transverse and symmetric tensor 
$\Pi^{\mu\nu}$,
\be
\label{equ:fluc-eom3}
\Pi^{\mu\nu}=-\gamma^{\mu\nu\alpha\beta}\frac{\nabla_{(\alpha} u_{\beta)}}{T}\,.
\ee
In addition to the symmetry between pairs of indices $(\mu\nu)$ and $(\alpha\beta)$, which
is guaranteed by Onsager's relation~\cite{Landau-sp2}, %it should be emphasized that 
$\gamma^{\mu\nu\alpha\beta}$ is symmetric, traceless (with conformal EOS) and transverse in 
$\mu$ and $\nu$, as well as in $\alpha$ and $\beta$.
It can be shown that for Navier-Stokes hydrodynamics $\Pi^{\mu\nu}=-\eta\sigma^{\mu\nu}$
\cite{Kapusta:2011gt}\footnote{
Except the result shown in \Eq{equ:fluc-correlation},
when $\zeta\ne 0$ 
there is an extra piece of $\gamma_\zeta^{\mu\nu\alpha\beta}$,
\be
\gamma^{\mu\nu\alpha\beta}_\zeta=2T\zeta \Delta_\zeta^{\mu\nu\alpha\beta}\,,
\ee
where $\Delta_\zeta^{\mu\nu\alpha\beta}=\frac{1}{2}\Delta^{\mu\nu}\Delta^{\alpha\beta}$
is factorized automatically in terms of the projection operator.},
\be
\label{equ:fluc-correlation}
\gamma^{\mu\nu\alpha\beta}=2T\eta\Delta^{\mu\nu\alpha\beta}\,,
\ee
with the tensor structure $\Delta^{\mu\nu\alpha\beta}$ defined by projection operators,
\be
\Delta^{\mu\nu\alpha\beta}
=\frac{1}{2}\left[\Delta^{\mu\alpha}\Delta^{\nu\beta}+\Delta^{\mu\beta}\Delta^{\nu\alpha}\right]
-\frac{1}{3}\Delta^{\mu\nu}\Delta^{\alpha\beta}\,.
\ee

The two-point auto-correlation determined by \Eqs{equ:fd-theorem} and 
(\ref{equ:fluc-eom3}) are characteristic only for the first order dissipative
hydrodynamics, which represents white noise with correlation strength
constrained by first order transport coefficients. %$\eta$. 
%Dirac delta function in realistic systems reduces 
%to the inverse of system size in space-time, 
%\[
%\delta^{(4)}(x_1-x_2)\sim \frac{1}{\Delta t \Delta V}\,,
%\]
%which na\"ively explains why hydrodynamical noise
%should be expected more pronounced in smaller systems. 
When higher order viscous corrections are taken into account
with respect to the causality, 
space-time dependence must be altered from a Dirac delta function
and one accordingly obtains colored noise which depends also 
on higher order transport coefficients~\cite{Murase:2013tma}.
Although Navier-Stokes hydrodynamics suffers from causality problem,
which contains superluminal modes corresponding to sufficiently large
momentum. In this work, we shall only solve Navier-Stokes hydrodynamics  
with finite number of hydrodynamic modes, in a way that the issue of causality
becomes less significant.

\section{Gubser flow}
\label{sec:gubser}

In this section we briefly review some of the essential ingredients of Gubser flow, which
are relevant to our present work.  
More details of Gubser's solution to relativistic hydrodynamics and discussions 
can be found in \cite{Gubser:2010ze,Gubser:2010ui}.
%, with essential ingredients relevant to this  work   

Analytical solutions to viscous hydrodynamics can be achieved with respect to %when hydrodynamics 
%is simplified under 
certain symmetry constraints, 
which allows one to recover the solution in a more general coordinate system
by coordinate transformations.
%Consequently, one is allowed to do a coordinate transformation which
%recovers the solution in a more general coordinate system.
For instance, the well-known Bjorken flow relies on a Bjorken boost invariance
regarding the space-time geometry in heavy-ion collisions, 
which is explicit if one writes the metric tensor 
in Milne coordinates (beam axis along $z$),
\be
\label{equ:milne}
ds^2=-d\tau^2 + dx^2 + dy^2 + \tau^2 d\xi^2,
\ee
with 
\[
\tau = \sqrt{t^2-z^2}\quad \mbox{and}
\quad
\xi=\frac{1}{2}\ln \frac{t+z}{t-z}\,,
\]
and one immediately finds the solution of flow velocity profile with $u^\xi=0$,
which then gives rise to $v_z=z/t$ in the original space-time.
Recently, Gubser and Yarom
developed a new type of analytical solution by imposing %additionally 
rotational symmetry with respect to the beam axis~\cite{Gubser:2010ze,Gubser:2010ui},
in addition to Bjorken boost invariance. Starting from
the Milne space-time \Eq{equ:milne}, 
the manifold $R^{3,1}$ is firstly transformed into $dS_3\times R$
via a Weyl rescaling. The following mapping, 
\begin{align}
\label{equ:mapping}
\sinh\rho=&-\frac{1-q^2\tau^2+q^2r^2}{2q\tau}\,,\\
\tan\theta=&\frac{2qr}{1+q^2\tau^2-q^2r^2}\,,
\end{align}
further allows one to rewrite the metric tensor as
\be
\label{equ:dshat}
d{\hat s}^2=-d\rho^2+d\xi^2
+\cosh^2\rho\left(d\theta^2+\sin^2\theta d\phi\right)\,,
\ee
which manifests the symmetry $SO(3)\times SO(1,1)\times {\mathcal Z}_2$. 
$\rho$ in \Eq{equ:dshat} is interpreted as the de Sitter time.
$q$ in \Eq{equ:mapping} is a parameter with dimension inverse of length, which
characterizes the transverse size of the system.
One can accordingly read off
the profile of flow velocity from \Eq{equ:dshat}
  
\be
\label{equ:cellrest}
\hat u^\mu=(1,0,0,0)\,.
\ee
We follow the same notation in \cite{Gubser:2010ui} that 
a `hat' over a quantity indicates that the quantity is defined 
in the coordinate system \Eq{equ:dshat}.
Although flow velocity profile in \Eq{equ:cellrest} 
describes a fluid cell at rest, there exists non-zero
expansion rate due to the geometry, 
\be
\label{equ:du}
\hat\nabla\cdot\hat u = 2 \tanh \rho\,,
\ee 
which results in a diagonal
shear tensor $\sigma^{\mu\nu}$,
\be
\label{eq:sigma_gub}
\hat\sigma^{\mu\nu} = \hat P^{\mu\nu} \frac{2}{3}\tanh \rho\,,
\ee
where $\hat P^{\mu\nu}$ is a \emph{traceless} projection operator which is 
identical to $\hat \Delta^{\mu\nu}$ except for the $\xi\xi$ component:
$\hat P^{\xi\xi}=-2\hat\Delta^{\xi\xi}$.
%which describes a fluid cell at rest. 
Simplification of hydro equations of motion is expected accordingly. 
And indeed the only non-trivial
equation left from \Eq{equ:hydro-eom} is an equation of continuity,
\be
\label{equ:hateom}
D\hat \e + (\hat \e+\hat \P) d_\mu \hat u^\mu
+\nabla_{(\mu}\hat u_{\nu)}\hat \Pi^{\mu\nu}=0\,,
\ee
which still requires solution with respect to the gradient expansion order by order.
%To the first order, gradients of flow velocity result in
%\be
%\label{equ:du}
%\hat\nabla\cdot\hat u = 2 \tanh \rho\,,
%\ee 
%and
%\be
%\hat\sigma^{\mu\nu} = \bar \Delta^{\mu\nu} \frac{2}{3}\tanh \rho\,,
%\ee
%where $\bar \Delta^{\mu\nu}$ is a \emph{traceless} projection operator which is 
%identical to $\hat \Delta^{\mu\nu}$ except the $\xi\xi$ component:
%$\bar \Delta^{\xi\xi}=-2\hat\Delta^{\xi\xi}$.  
The analytical solution to Navier-Stokes hydrodynamics can be found for a conformal
equation of state: $\hat\e=3\hat\P$. For instance, 
the energy density in the `hat' system is~\cite{Gubser:2010ui}
\be
\hat \e(\rho)=(\cosh\rho)^{-\frac{8}{3}}\left[\hat T_0+\frac{1}{3}H_0F(\rho)\right]^4\,,
\ee
where $F(\rho)$ is a function defined by the following integral
\be
F(\rho)=\int_0^\rho dr( \cosh(r))^{\frac{2}{3}}\tanh^2(r)\,,
\ee
and $H_0$ is a constant proportional to $\eta/s$ which parameterizes shear 
viscosity in the `hat' coordinate system. $\hat T_0$ is a dimensionless 
parameter constrained by the total multiplicity of the collision event,
whose details will be given later in \Sect{sec:solution} regarding numerical simulations.

Once hydrodynamics is fully solved in the `hat' coordinate system, quantities in the $(\tau,x,y,\xi)$ 
system can be recovered via the following transformations,
\begin{subequations}
\begin{align}
\label{equ:trans-law}
\epsilon &= \tau^{-4}\hat \epsilon\,,\\
n &= \tau^{-3}\hat n\,,\\
u_\tau &= \tau\left(\frac{\partial\rho}{\partial\tau}\hat u_\rho+\frac{\partial\theta}{\partial\tau}\hat u_\theta\right)\,,\\
u_\perp&= \tau\left(\frac{\partial\rho}{\partial r}\hat u_\rho+\frac{\partial\theta}{\partial r}\hat u_{\theta}\right)\,,\\
u_{\phi_i}&= \tau\hat u_{\phi_i}\,,\\
u_{\xi}&=\tau\hat u_{\xi}\,.
\end{align}  
\end{subequations}

It has been noticed that the gradient expansion breaks down in Gubser's solution
at early de Sitter times for Navier-Stokes hydrodynamics~\cite{Gubser:2010ui},
and also at late de Sitter times if one investigates second order viscous hydrodynamics. 
Such constraints indicate initialization times at which the thermal system can be
approximated by viscous hydrodynamics. Accordingly, regarding realistic systems
in heavy-ion collisions, we shall avoid very early or late de Sitter times in  
our analysis. 
%to the regime which corresponds
%to a initial thermalization time of the order of $1$ fm/c.

\section{Noisy Gubser flow}
\label{sec:noisy-gubser}

%Symmetries not only simplify hydro equations of motion, but also the structure of
%two-point autocorrelations of hydrodynamical noise. 

To a large extent, effects of hydrodynamical noise can be investigated analytically 
with respect to solvable hydro models, such as Bjorken flow in 1+1D and Gubser flow. 
The discussion on hydrodynamical noise with respect to Bjorken flow in 1+1D
was given previously in \cite{Kapusta:2011gt}, from which we generalize to the case
of Gubser flow. 

\subsection{Bjorken flow}
\label{sec:bjor}

Following the discussions in \cite{Kapusta:2011gt}, 
for a 1+1D Bjorken's solution with respect to the
Navier-Stokes hydrodynamics, the essential simplification is
a factorization of the tensor structure of the two-point autocorrelation 
function in \Eq{equ:fluc-correlation}
\be
\Delta^{\mu\nu\alpha\beta}\propto P^{\mu\nu} P^{\alpha\beta}\,,
\ee
and one thus expects 
\be
\label{equ:bjork-S}
S^{\mu\nu}(\tau,\xi) = w(\tau)f(\tau,\xi)P^{\mu\nu}\,, 
\ee
where $P^{\mu\nu}$ is the same tensor defined under \Eq{eq:sigma_gub}, but now in the 
Milne space-time.
Notice that we write the factorization in \Eq{equ:bjork-S} 
in terms of the traceless projector $P^{\mu\nu}$ 
instead of $\Delta^{\mu\nu}$~\cite{Kapusta:2011gt}, 
which differs 
by a factor of two in this particular case of Bjorken flow in 1+1D.\footnote{
Since the bulk part is automatically factorized, one can simply restore the result in 
\cite{Kapusta:2011gt} for the $\zeta\ne 0$ case with an extra term 
proportional to $\Delta^{\alpha\beta}$
}
Generally, for a conformal fluid system, $P^{\mu\nu}$ is preferred 
since $\gamma^{\mu\nu\alpha\beta}$ is traceless in $\mu$ and $\nu$.
With dimension being saturated by the 
enthalpy density $w(\tau)$ in \Eq{equ:bjork-S} and tensor structure 
being represented by $P^{\mu\nu}$, 
hydrodynamical noise associated with shear viscous tensor 
is reduced to one unknown scalar function $f(\tau,\xi)$, which satisfies
\be
\label{equ:bjo-cor}
\bra f(\tau_1,\xi_1) f(\tau_2,\xi_2) \ket=\frac{2T(\tau_1)\eta(\tau_1)}{3A_\perp w^2(\tau_1) \tau_1}\delta(\tau_1-\tau_2)
\delta(\xi_1-\xi_2)\,.
\ee
$A_\perp$ characterizes transverse size of the system. 

Noisy hydro equations of motion %in the case of thermal fluctuations 
can be simplified with respect to the factorization. 
For later convenience, we expand the $\xi$ dependence of a hydrodynamical
variable into its
conjugate $\kx$, 
through a Fourier transformation, e.g., temperature is
\be
T(\tau,\xi)=\int \frac{d\kx}{2\pi}e^{i\kx\xi} \t T(\tau,\kx)\,.
\ee
One finds that for each $\kx$-mode, \Eqs{equ:fluc-eom2} can now be recast 
into a Langevin equation,
\be
\label{equ:bjo-fluc-eom}
\t\V'(\rho,\kx)=-\t\Gamma(\rho,\kx)\t\V(\rho,\kx) +\t \K(\rho,\kx)\,,
\ee
where prime indicates derivatives with respect to\footnote{
We use the same notation $\rho$, which should be distinguished from the 
de Sitter time $\rho$ used in the %coordinate in the 
`hat' coordinate system for Gubser flow. 
%although they both can be interpreted as time.  
} $\rho=\ln(\tau/\tau_0)$
and
\be
\t\V(\rho)=
 \begin{pmatrix}
  \t n(\rho,\kx) \\
  \t\alpha(\rho,\kx)
 \end{pmatrix}\,.
\ee
We follow the same notations as in~\cite{Kapusta:2011gt}, such that
$\t n=\int d\xi e^{i\kx\xi}\delta s/s$ stands for the 
relative fluctuations
of entropy density, and $\t \alpha=\int d\xi e^{i\kx\xi} \tau u^\xi$ the fluctuation 
of flow velocity along $\xi$. 
$\t\Gamma$ is a $2\times2$ 
matrix which is entirely determined by Bjorken's solution of hydrodynamics~\cite{Kapusta:2011gt}. 
$\t \K$ incorporates $\kx$-mode of the random 
scalar function $f(\tau,\xi)$, 
\be
\t\K(\rho)
= \begin{pmatrix}
  -\t f \\
  -ik_\xi\t f 
 \end{pmatrix}\,.
\ee
It is interesting to note that the effect of thermal noise on $\t\alpha$ vanishes for the $\kx=0$ mode.

Without transverse expansion, Bjorken flow in 1+1D is an oversimplified model regarding heavy-ion collisions.
Nonetheless, it is worth analyzing, at least to a qualitative level, the effect of hydrodynamical
noise. We start by rewriting \Eq{equ:bjo-cor} for each $\kx$-mode as,
\be
\label{equ:bjo-2p}
\bra \t f(\tau_1,\kx_1) \t f(\tau_2,\kx_2) \ket
=\frac{\pi \nu}{A_\perp w(\tau_1) \tau_1}\delta(\tau_1-\tau_2)
\delta(\kx_2+\kx_2)\,.
\ee  
Despite the constant $\nu$, which is proportional to $\eta/s$,
\[
\nu = \frac{4}{3}\frac{\eta}{s}
\] 
the amplitude of the autocorrelation is 
totally determined by the factor $A_\perp w \tau$ in the denominator.
Apparently one would expect a stronger effect of hydrodynamical noise in
a system with smaller size, due to the appearance of $A_\perp$.
However, with respect to heavy-ion collisions, parametrically
one can write the factor as a whole as 
\be
A_\perp w(\tau) \tau\sim A_\perp \left(\frac{dE}{\tau d^2 x_\perp d\xi}\right) \tau \sim 
\frac{dE_\perp}{d y}\,,
\ee
which is interpreted as the transverse energy deposited per rapidity, and 
is equivalent to the multiplicity production of one collision event.
\Eq{equ:bjo-2p} thus demonstrates the fact that the strength of hydrodynamical noise
is essentially controlled by the multiplicity, instead of transverse size of 
a colliding system.

\subsection{Gubser flow}

Following the same strategy for Bjorken flow in the previous section, 
we generalize the discussion of hydrodynamical noise to the case of Gubser flow.
Note that in the $(\rho,\theta,\phi,\xi)$ coordinate system, 
flow velocity profile is essentially the same as a 
Bjorken's solution in 1+1D, except the fact that 
transverse expansion is now taken into account as well. 
Therefore, it is not surprising that the
symmetry considered in Gubser flow leads to a similar factorization of the 
tensor structure in \Eq{equ:fluc-correlation}.
Especially for Navier-Stokes hydrodynamics 
$\hat\Pi^{\mu\nu}\propto \hat P^{\mu\nu}$,
one has
\be
\hat\Delta^{\mu\nu\alpha\beta}\propto \hat P^{\mu\nu}\hat P^{\alpha\beta}
\ee
and thus the autocorrelation of hydrodynamical noise becomes 
\be
\bra\hat S^{\mu\nu}(\rho_1,\theta_1, \phi_1,\xi_1)\hat S^{\alpha\beta}(\rho_2,\theta_2,\phi_2,\xi_2)\ket
=\frac{\nu \hat T\hat s  \hat P^{\mu\nu}\hat P^{\alpha\beta}}{2\cosh^2\rho_1\sin\theta_1}
\delta(\rho_1-\rho_2)\delta(\theta_1-\theta_2)\delta(\phi_1-\phi_2)\delta(\xi_1-\xi_2)\,.\\
\ee
With the above factorization, we 
can accordingly write $\hat S^{\mu\nu}$ in terms of $\hat P^{\mu\nu}$,
\be
\label{equ:gub-factor}
\hat S^{\mu\nu}(\rho,\theta,\phi,\xi)=\hat w(\rho) \hat f(\rho,\theta,\phi,\xi)\hat P^{\mu\nu}\,,
\ee
so that the correlation function of hydrodynamical noise 
reduces to correlation of a dimensionless scalar function,
\be
\bra \hat f(\rho_1,\theta_1, \phi_1,\xi_1)\hat f(\rho_2,\theta_2,\phi_2,\xi_2)\ket
=\frac{\nu }{2\hat w \cosh^2\rho_1 \sin\theta_1}
\delta(\rho_1-\rho_2)\delta(\theta_1-\theta_2)\delta(\phi_1-\phi_2)\delta(\xi_1-\xi_2)\,.
\ee
Note that enthalpy density $\hat w(\rho)$ is used to saturate conformal dimension, even though
quantities in the `hat' system are dimensionless by construction. 

Accounting for
the $SO(3)\times SO(1,1)\times \mathcal{Z}_2$-symmetry of Gubser flow, 
the mode decomposition
can be done with respect to spherical harmonics. In particular for the scalar 
function $\hat f(\rho,\theta,\phi,\xi)$, one has
\footnote{
Our convention of spherical harmonics is 
\[
Y_{lm}(\theta,\phi)=\sqrt{\frac{2l+1}{4\pi}\frac{(l-m)!}{(l+m)!}}P_l^m(\cos(\theta)) e^{im\phi}.
\]
}
\be
\hat f(\rho, \theta, \phi, \xi) = \sum_{l,m} \int \frac{d \kx}{2\pi}
h_{lm}(\rho,\kx)Y_{lm}(\theta, \phi)e^{ik_\xi \xi}\,.
\ee
and the two-point autocorrelation of each mode can be correspondingly found as
\be
\label{equ:mode-twop}
\bra h_{l_1m_1}(\rho_1,\kx_1)h_{l_2m_2}(\rho_2,\kx_2)\ket
=\frac{\pi\nu}{\hat w \cosh^2\rho_1}\delta(\rho_1-\rho_2)\delta_{l_1l_2}
\delta_{m_1,-m_2}(-1)^{m_1}\delta(\kx_1+\kx_2)\,.
\ee
%Note that in the above correlation there is no contributions from vector modes. 
A couple of comments are in order with respect to \Eq{equ:mode-twop}.
First, it should be emphasized that hydrodynamical noise contains only scalar modes, 
as a direct consequence of the factorization \Eq{equ:gub-factor}. 
However, it is not generally true since the tensor structure of the hydrodynamical noise
term can be more involved and leads to fluctuations in both vector and tensor modes.
Second, one can parametrically estimate the effects of 
hydrodynamical noise in heavy-ion collisions,
by examining the magnitude of correlation in \Eq{equ:mode-twop}, as we did previously
in \Sect{sec:bjor} for the Bjorken flow. The essential
quantity that determines the strength of hydrodynamical noise is $\hat w$ in the 
denominator, in addition to the constant parameter $\nu$ which is proportional to $\eta/s$. 
In Gubser's solution to hydrodynamics, $\hat w$ relies solely on the parameter 
$\hat T_0$, which is determined by the total multiplicity, but \emph{not system size}. 
Therefore, in accordance with what was noticed in \Sect{sec:bjor},
although one usually expects the system size to play a significant role in
the estimate of hydrodynamical noise, we conclude that
the absolute effect of hydrodynamical noise in a expanding medium in heavy-ion 
collisions is dominated by \emph{multiplicity}. 

When decomposing fluctuations of hydro variables, scalar modes and vector modes must be considered
with respect to spherical harmonics $Y_{lm}$ and vector spherical harmonics $\Phi_{lm}^i$ 
respectively. We expand the following independent fluctuation variables,
\begin{subequations}
\label{equ:mode-decom}
\begin{align}
\delta \hat T(\rho,\theta,\phi,\xi) =& \hat T(\rho) \sum_{l,m} \int \frac{d \kx}{2\pi}
\delta_{lm}(\rho,\kx) Y_{lm}(\theta,\phi) e^{ik_\xi \xi}\,,\\
\delta u_i(\rho,\theta,\phi,\xi) =& \sum_{l,m} \int \frac{d \kx}{2\pi}\left[v_{lm}^s(\rho,\kx)
\partial_i Y_{lm}(\theta,\phi) 
+ v_{lm}^v(\rho,\kx) \Phi_{lm}^i(\theta, \phi)\right]e^{ik_\xi \xi}\,,\\
\delta u_\xi(\rho,\theta,\phi,\xi) =& \sum_{l,m} \int \frac{d \kx}{2\pi}v_{lm}^\xi(\rho,\kx) 
Y_{lm}(\theta,\phi)e^{ik_\xi \xi}\,,
\end{align}
\end{subequations}
where $i=\theta, \phi$ denotes orientations in the transverse plane. 
In terms of all the scalar modes ($\delta_{lm}$,
$v^s_{lm}$, $v^\xi_{lm}$) and vector modes $v^v_{lm}$, 
\Eqs{equ:fluc-eom2} on top of Gubser flow take an identical form as 
in \Eq{equ:bjo-fluc-eom},  
\be
\label{equ:gub-fluc-eom}
\t\V_{lm}'(\rho,\kx)=-\t\Gamma(l,m,\rho,\kx)\t\V_{lm}(\rho,\kx) +\t \K_{lm}(\rho,\kx)\,,
\ee 
with extra dependence on the transverse dimension captured 
by indices $l$ and $m$. The prime in \Eq{equ:gub-fluc-eom}
indicates derivative with respect to $\rho$, and $\t\V(\rho)$ is,
\be
\t\V_{lm}(\rho,\kx)=
 \begin{pmatrix}
  \delta_{lm}(\rho,\kx) \\
  v_{lm}^s(\rho,\kx) \\
  v_{lm}^\xi(\rho,\kx)  \\
  v_{lm}^v(\rho,\kx)
 \end{pmatrix}\,.
\ee 
We purposely assign the vector mode to be the fourth element in $\t\V_{lm}$.
$\t\Gamma(l,m,\rho,\kx)$ in \Eq{equ:gub-fluc-eom} is a $4\times 4$ matrix with 
a rather complicated from, with its components given in~\cite{Gubser:2010ui}. 
%and \App{}. 
$\t\Gamma$ is block-diagonalized, since
vector modes and scalar modes are decoupled due to parity. %symmetry arguments.
$\t\K_{lm}(\rho,\kx)$ depends
on the scalar modes of hydrodynamical noise $h_{lm}$,  
\be
\label{eq:noise-gub}
\t\K_{lm}(\rho,\kx)
= \begin{pmatrix}
  -\frac{2}{3}\tanh\rho \,h_{lm}(\rho,\kx) \\
  \frac{2\hat T}{3\hat T'}\tanh\rho \,h_{lm}(\rho,\kx) \\
  \frac{i4k_\xi \hat T}{\hat T+H_0\tanh \rho} h_{lm}(\rho,\kx)  \\
 0
 \end{pmatrix}\,.
\ee 
The last element in $\t\K$ vanishes as it should, since hydrodynamical noise does not 
contribute to vector modes. %For the similar reason, $\t\Gamma$ is block-diagnolized.
Also, we notice that for the $\kx=0$ mode, the evolution of $v^\xi_{lm}(\rho, 0)$ 
is insensitive to the hydrodynamical noise as well.  
\Eq{eq:noise-gub} introduces two sources of instabilities, corresponding to the 
zeros of $\hat T'$ and $\hat T + H_0 \tanh\rho$. As has been discussed 
throughly in \cite{Gubser:2010ui}, both sorts of instabilities %are not new which 
have been noticed in the structure of the matrix $\t\Gamma$
already in the case without hydrodynamical noise.
%which however are not
%significant at the de Sitter times that are relevant to a realistic 
%heavy-ion system.
%can be avoided in the analysis at de Sitter times that are not too early.  

\section{Solving Gubser flow with hydrodynamical noise}
\label{sec:solution}

Hydro equations of motion %\Eq{equ:hydro-eom} 
are coupled with the equation 
of state, which in general resorts to
numerical solutions. %When being linearized regarding small perturbations 
%due to fluctuations, 
When hydrodynamical noise is taken in account, 
in addition to the ensemble-averaged background flow, 
one also needs numerical simulations of stochastic equations~\cite{Young:2013fka,Young:2014pka}
regarding the random unknown tensor variables $S^{\mu\nu}$. 
%Instead of explicit investigation of numerical simulations of noisy hydrodynamics, 
%we restrict ourselves to the case of Gubser flow.
However, with respect to a background Gubser flow, the tensor structure 
of hydrodynamical noise is well determined
and has been discussed in the previous section,
%, tensor structure of hydrodynamical
%noise is well constrained with respect to Gubser flow, which 
so that numerical simulations are largely simplified. 

A general procedure of solving noisy hydrodynamics on top of Gubser flow comprises the 
following steps. First, one solves \Eqs{equ:gub-fluc-eom} mode-by-mode, with 
initial conditions and parameters specified with respect to desired collision systems.
Second, hydrodynamical variables, such as flow velocity and temperature, are 
obtained through mode summation as in \Eqs{equ:mode-decom}. 
Similar strategy has also been applied in analyses of perturbations on top of solvable hydro
models~\cite{Florchinger:2011qf,Brouzakis:2014gka,Staig:2010pn}.
To recover quantities in the original
$(\tau,r,\phi,\xi)$ coordinate system, one is additionally required to do a coordinate 
transformation according to \Eqs{equ:mapping}.

%\subsubsection{Numerical solution of stochastic equations}

For each mode, although \Eqs{equ:gub-fluc-eom} represent four-coupled equations, 
simplifications can be made as follows. %regarding the following facts. 
First, 
as we emphasized before, the hydrodynamical evolution 
does not couple vector modes to scalar modes. Besides, hydrodynamical 
fluctuations do not contribute to vector modes. As a consequence, we shall not
consider vector modes in our analysis. 
%which has already been discussed thoroughly in \cite{Gubser:2010ui}. 
Second, we shall restrict ourselves to the case
of $\kx=0$. By doing so, the matrix $\t\Gamma$ is further block-diagonalized in
a way that equation of motion for $v^\xi$ is decoupled. 
The third component of $\t\K$ also vanishes when $\kx=0$.
%namely, hydrodynamical fluctuations do not affect the evolution of $\kx=0$ mode of $v^\xi$.
Therefore, the only non-trivial equations of motion which receive 
extra contributions from hydrodynamical noise are the two coupled equations 
for the scalar modes $\delta_{lm}$ and $v^s_{lm}$, with  
%\be
%\t\Gamma(l,m,\rho,\kx)=
%\begin{pmatrix}
%  \frac{H_0\tanh^2\rho}{3\hat T} & \frac{l(l+1)\sech^2\rho[H_0\tanh\rho-\hat T]}{3\hat T}\\
%  \frac{2H_0\tanh\rho}{H_0\tanh\rho-2\hat T}+1  & 
%\frac{H_0\hat T[-4(3l(l+1)-10)\sech^2\rho-16]+6H_0^2\tanh^3\rho+8\hat T^2\tanh\rho}
%{6\hat T[H_0\tanh\rho-2\hat T]}
% \end{pmatrix}\,,
%\ee
\begin{subequations}
\label{equ:gammas}
\begin{align}
\t\Gamma_{11}=&\frac{H_0\tanh^2\rho}{3\hat T}\,\\
\t\Gamma_{12}=&\frac{l(l+1)\sech^2\rho[H_0\tanh\rho-\hat T]}{3\hat T}\,\\
\t\Gamma_{21}=&\frac{2H_0\tanh\rho}{H_0\tanh\rho-2\hat T}+1\,\\
\t\Gamma_{22}=&\frac{H_0\hat T[-4(3l(l+1)-10)\sech^2\rho-16]+6H_0^2\tanh^3\rho+8\hat T^2\tanh\rho}
{6\hat T[H_0\tanh\rho-2\hat T]}
\end{align}
\end{subequations}
Similar equations have been investigated in \cite{Staig:2010pn} without hydrodynamical noise.
Note that all $m$-modes of the same index $l$ evolve identically since there is 
no dependence on index $m$ in the $\t\Gamma$ matrix and $\t\K$. 
%thus we can suppress index $m$ in what following for simplicity.

\subsection{Formal solution}
\label{sec:formal-solution}

Before numerically solving \Eqs{equ:gub-fluc-eom}, we investigate qualitatively the
behavior of hydrodynamical noise and its evolution. 
The following discussion is made ad hoc with respect to a background Gubser flow,
but it can be applied to hydrodynamical noise on top of Bjorken flow as well.
One can write a formal solution of \Eq{equ:gub-fluc-eom} (and similarly \Eq{equ:bjo-fluc-eom})
in terms of a Green function,
%Formally, solution with respect to \Eq{equ:gub-fluc-eom} 
%(also for \Eq{equ:bjo-fluc-eom}) 
%can be found through a Green's function 
\be
\label{eq:1point}
\t\V(\rho, K) = \int_{\rho_0}^\rho d \rho' \t\G(\rho-\rho', K)\t\K(\rho', K)+\t\G(\rho-\rho_0,K)\t\V(\rho_0,K),
\ee
where $K$ is an abbreviated notation for the conjugate variables to %for $(l,m,\kx)$ 
specify modes, i.e., $K=(l,m,\kx)$ for Gubser flow and
$K=\kx$ for Bjorken flow.
Green function $\t\G$ is determined by the following equation 
(note that we take $\rho=\ln(\tau_0/\tau)$ for Bjorken flow)
\be
\label{eq:green0}
\partial_\rho \t\G(\rho-\rho',K) = -\t\Gamma(\rho,K) \t\G(\rho-\rho',K)\,,
\ee
with the initial condition 
\be
\t\G(0, K)=1\,.
\ee
The solution of \Eq{eq:green0} is
\be
\label{equ:green1}
\t\G(\rho-\rho',K)=\mathcal{T}\exp\left[-\int_{\rho'}^\rho d\rho''\t\Gamma(\rho'',K)\right]\,,
\ee
where $\mathcal{T}$ indicates a time ordering with respect to $\rho$.
Except in some extreme limits, there is no simple analytical 
expression for the Green function $\t\G$, regarding 
a specified form of $\t\Gamma$ from hydrodynamics. However, some of 
the behaviors of mode evolution are known qualitatively. For instance,
it has been shown that viscosity damps mode evolution~\cite{Gubser:2010ui}.
In particular,
in the large $l$ and small viscosity limit, one finds that 
$\t\G(\Delta\rho)\sim \exp[-l^2 H_0\Delta \rho]$.
%The above discussion from \Eqs{eq:1point} to (\ref{equ:green1}) can also be applied to 
%hydrodynamical noise on top of Bjorken flow in 1+1D, with $\rho=\ln(\tau_0/\tau)$ and
%$K=\kx$.  

From \Eq{eq:1point}, we notice that at any time, 
each mode comprises contributions from fluctuations in 
the initial state (the second term) and 
hydrodynamical noise (the first term).
%One-point function of the fluctuations follows \Eq{eq:1point}. 
Especially, due to the fact that the ensemble average of one-point
function of thermal fluctuations vanishes, 
\[
\bra \t\K\ket = 0\,,
\]
the mode evolution of the one-point function is 
governed by hydrodynamic response to initial state
fluctuations. In heavy-ion collisions, initial state fluctuations are %understood
%as a consequence of quantum effect, which is 
quantum fluctuations associated with the probability distribution
%wave function
of nucleons inside the colliding nucleus. %on an event-by-event basis
Therefore, one would expect initial state fluctuations to be independent of
ensemble average, and the one-point function evolution has a
form of linear response
%\footnote{
%In principle, one should be able to further relate $\t G(\rho, K)$ to the two-point correlation of conserved currents.
%For instance, for fluctuations of transverse flow velocity, $\t G[\delta u_i]$ can be obtained by 
%\[
%\t G^{0i0i}_{R}=i\theta(t) \bra[T^{0i},T^{0i}]\ket
%\]
%}
\be
\bra \t \V(\rho,K)\ket = \t\G(\rho-\rho_0,K)\bra \t\V(\rho_0,K)\ket.
\ee
%which can be seen as the general hydro EoM describing the evolution of initial-state fluctuations in one event. 

The effect of hydrodynamical noise can be 
investigated in terms of the evolution of
two-point correlation function.
%Taking ensemble average with respect to the square of \Eq{eq:1point},  
The equal-time two-point correlation can be found according to \Eq{eq:1point}
%
%The two-point correlation function of (thermal) fluctuations can be accordingly determined. With capital $X$ indicating 
%coordinates other than $\rho$, the equal-time two-point correlation function is
%\begin{align}
%\label{eq:2point}
%\bra \V_i(\rho,X_1)\V_j(\rho,X_2)\ket=&\int_{\rho_0}^\rho d\rho'\int d^{D-1}X \left(\G(\rho-\rho',X_1-X)
%\Lambda_{th}(\rho',X)\G^{T}(\rho-\rho',X_2-X)\right)
%\nonumber\\
%&+\int d^{D-1}X\left(\G(\rho-\rho_0,X_1-X)\Lambda_{ini}(\rho_0,X)\G^{T}(\rho-\rho_0,X_2-X)\right)\,,
%\end{align}
%which corresponds to the two-point function with respect to mode-$K$ (conjugate to $X$)
\begin{align}
\label{eq:2point-k}
\bra \t\V_i(\rho, K)\t\V_j(\rho, K')\ket=&
\int_{\rho_o}^{\rho} d\rho'\left(\t\G(\rho-\rho', K)\Lambda_{th}
(\rho')\t \G^T(\rho-\rho', -K)\right)_{ij}\delta(K+K')\nonumber\\
&+\left(\t\G(\rho-\rho_0, K)\Lambda_{ini}(\rho_0,K, K')\t \G^T(\rho-\rho_0,K')\right)_{ij}\,,%\delta(K+K')\,,
\end{align}
where subscript $i$ indicates the i-th component of the fluctuation mode $\t\V$,
and $\delta(K+K')$ stands for
\[
\delta_{l_1l_2}
\delta_{m_1,-m_2}(-1)^{m_1}\delta(\kx_1+\kx_2)\,,
\]
corresponding to the correlation of the white hydrodynamical noise on top of Gubser flow.
In obtaining \Eq{eq:2point-k}, %we have made several assumptions. First, 
we have extended the concept of `average'. %with respect to a more realistic modeling of heavy-ion collisions. 
We restrict ourselves to one specified
centrality class with fixed total multiplicity, so that the average in \Eq{eq:2point-k}
automatically contains average over collision events and ensemble average 
in each collision event,
\be
\bra\ldots\ket = \frac{1}{N_{\mbox{\small collision event}}}\sum 
\bigg[\bra\ldots\ket_{\mbox{\small ensemble}} \bigg]\,.
\ee
More explicitly, collision events are distinguished by their initial conditions, while
ensemble events stick with the same initial condition but evolve with random
hydrodynamical noise on an event-by-event basis.
Therefore,
we are allowed to drop the mixing between hydrodynamical noise $\t\K$ at any time
and initial state fluctuations $\t\V(\rho_0,K)$ in the average, namely
$\bra \t\K(\rho,K) \t\V(\rho_0,K')\ket=0$, which factorizes into the ensemble average
of one-point function of hydrodynamical noise, and
the two-point correlation in \Eq{eq:2point-k} is written again as a term from initial
state fluctuations plus a term from hydrodynamical noise.

The structure of two-point 
autocorrelations of hydrodynamical noise is known locally in space-time, with
the amplitude $\Lambda_{th}$ 
fixed with respect to the form of $\t\K$ given in \Eq{eq:noise-gub}
and the fluctuation-dissipation theorem in \Eq{equ:mode-twop}.
%\begin{align}
%\Lambda_{th}(\rho)
%%=&\frac{2(2\pi)\nu}{A_\perp w e^{2\rho}}
%%\begin{pmatrix}
%%  1 & ik_\xi\\
%%  ik_\xi  & -k_\xi^2
%% \end{pmatrix}
%%\quad\leftrightarrow\quad \mbox{Bjorken}\\
%=&\frac{\pi \nu}{\hat w \cosh^2\rho}
%\begin{pmatrix}
%  \frac{4}{9}\tanh^2\rho & -\frac{4\hat T}{9\hat T'}\tanh^2\rho & \frac{2ik_\xi \hat T\tanh\rho}{3(\hat T+H_0\tanh\rho)}\\
%   -\frac{4\hat T}{9\hat T'}\tanh^2\rho & \frac{4\hat T^2}{9\hat T^{'2}}\tanh^2\rho& \frac{2ik_\xi \hat T\tanh\rho}{3\hat T'(\hat T+H_0\tanh\rho)}\\
%\frac{2ik_\xi \hat T\tanh\rho}{3(\hat T+H_0\tanh\rho)}&\frac{2ik_\xi \hat T\tanh\rho}{3\hat T'(\hat T+H_0\tanh\rho)}
%&-\frac{k_\xi^2\hat T^2}{(\hat T+H_0\tanh\rho)^2}\\
% \end{pmatrix}
%%\quad\leftrightarrow\quad \mbox{Gubser}
%\end{align}
Note that vector modes do not contribute. Two-point correlations of initial state fluctuations
are determined by averaging over collision events, instead of ensemble average. 
We consider two extreme scenarios in this work. One is inspired
by independent sources~\cite{Blaizot:2014wba}, from which
two-point correlations of initial state fluctuations
are expected to be local in the transverse plane, %on an event-by-event basis, 
and all $K$-modes are initialized correspondingly with specified values. 
In the other scenario, we initialize with selected %some of the particular 
modes accounting for the $SO(2)$ rotational symmetry in the transverse
plane. Especially, we are allowed to deform initial energy density profile
with a desired eccentricity. 
%Strength of initial state fluctuations $\Lambda_{ini}$ can be consequently determined by the 
%magnitudes of azimuthal anisotropies based on phenomenology. 
More details of these two types of initialization will
be given later in the next section.
%related to the number of
%independent sources $N_s$, which is further related to the number of participant nucleons 
%in heavy-ion collisions.  
%For instance, it is known that initial entropy density fluctuates
%event-by-event, with 
%\[
%\frac{\delta s}{s}\simeq \frac{1}{\sqrt{N_s}}
%\]
%Therefore, neglecting fluctuations of flow velocity in the initial state 
%we arrive at
%\be
%\label{equ:ic}
%\Lambda_{ini}=\frac{\tau_0^2 A_\perp}{9N_s}\mbox{diag}(1,0,0,0)\,,
%\ee
%where the factor of 1/9 originates from the dimension difference between entropy density and temperature.

\subsection{Numerical simulations}
\label{sec:num-solution}
%We shall solve \Eqs{equ:gub-fluc-eom} on top of the known Gubser's solution of Navier-Stokes
%hydro. numerically. 

%Details of numerics of this work are given in \App{}.

%\subsubsection{Hydrodynamical noise in ultra-central PbPb, pPb and pp}

\begin{figure}
\begin{center}
\includegraphics[width=1.0\textwidth]{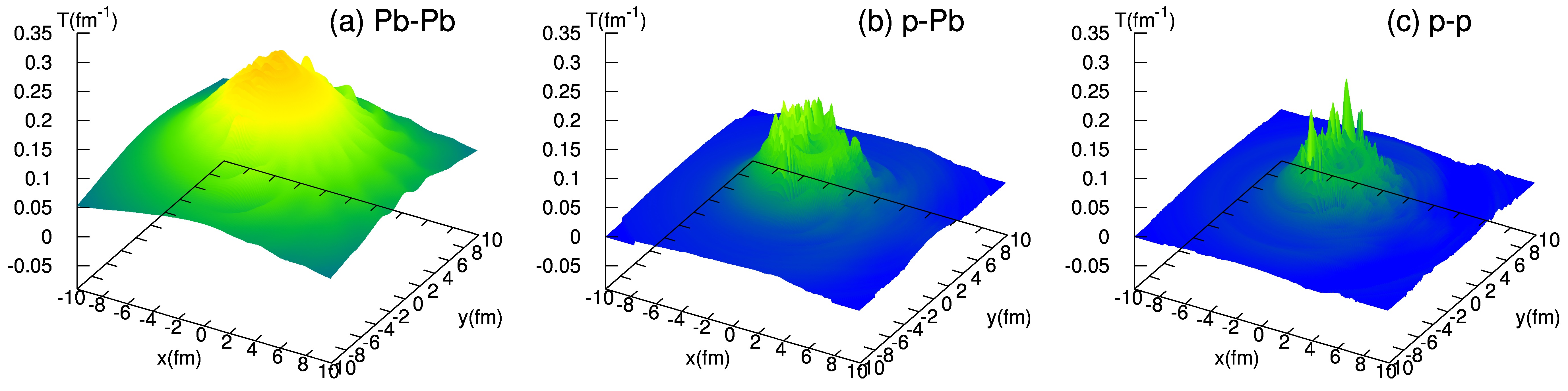}
\caption{\label{fig:snapshot}
(Color online)
Snapshots of temperature evolution at $\tau=2.5$ fm of one random 
(a) Pb-Pb, (b) p-Pb and (c) p-p event, 
in units of fm$^{-1}$.
}
\end{center}
\end{figure}

We solve the two coupled equations of motion numerically on top of Gubser 
flow for the scalar modes $\delta_{l}$ and $v^s_{l}$, 
with parameters $\hat T_0$ and 
$q$ specified with respect to ultra-central collision events 
of Pb-Pb, p-Pb and p-p at the LHC energies. 
$\hat T_0$ is determined according to the event multiplicity. Following
\cite{Gubser:2010ze} and \cite{Staig:2010pn}, we adopt the relation, 
\be
\hat T_0=\frac{1}{f_*^{1/12}}\left(\frac{3}{16\pi}\frac{dS}{d\xi}\right)^{1/3}\,,
\ee
where the constant $f_*=\epsilon/T^4=11$ is the effective degree of freedom 
extracted from Lattice QCD calculations, and
\be
\frac{dS}{d\xi}=7.5 \frac{dN_{ch}}{d y}\,.
\ee
$\hat T_0$ is found to be 7.3 for the 0$\sim$5\% central PbPb 
collisions with $\sqrt{s_{NN}}=2.76$ TeV~\cite{Staig:2010pn},
corresponding to multiplicity production per rapidity $dN_{ch}/dy\sim1600$~\cite{Aamodt:2010cz}. 
For the p-Pb collisions %carried out at the LHC 
with $\sqrt{s_{NN}}=5.02$ TeV~\cite{Chatrchyan:2013nka}, 
we take $\hat T_0=3.1$ which leads to an estimate of $dN_{ch}/dy\sim150$.  
Very recently, long-range correlations were measured in the 
ultra-central proton-proton collision events 
with multiplicity higher than 100 in the rapidity 
gap $|y|<2.4$~\cite{CMS:2015zpa}, for which we take $\hat T_0=2.0$.
The parameter $q$ constrains the finite transverse size of the fluid system.
For a Pb-Pb system,
we take $q=(4.3\mbox{fm})^{-1}$~\cite{Gubser:2010ze}, 
while for both p-Pb and p-p systems we take  
$q=(1.1\mbox{fm})^{-1}$. In all the simulations in this work, 
the viscosity parameter is taken to be 
$H_0=0.33$~\cite{Gubser:2010ui}, which corresponds to $\eta/s=0.134$. 
%Hydrodynamical noise results in random 
%fluctuations of temperature and flow velocity in space-time, which in turn affect
%medium evolution, as shown in \Fig{fig:snapshot}. 

\Fig{fig:snapshot} displays the solved temperature distribution 
%in the transverse plane 
of one random Pb-Pb, p-Pb and p-p event 
at $\tau=2.5$ fm, without initial state fluctuations. 
%Note that no initial state fluctuation 
%is introduced in these simulations.
%Comparing to the evolution of background Gubser flow (left panels), 
Specifying $\tau=2.5$ fm for the analysis of all the three systems breaks 
conformal symmetry, which is however of phenomenological 
interest since $\tau=2.5$ fm is a typical time scale that all
the three systems experience in the early stages of evolution.    
The effect of hydrodynamical noise is qualitatively 
captured by the bumpiness of the temperature profile,
which presents a clear trend of becoming more pronounced from the Pb-Pb
system (\Fig{fig:snapshot}(a)) to the p-p system (\Fig{fig:snapshot}(c)). 
%becomes more and
%more pronounced from the Pb-Pb system, to p-Pb and p-p system.
We have checked that when taking ensemble average, %effect of hydrodynamical noise 
the bumpiness
disappears in the temperature evolution, which corresponds to
$\bra\delta_l(\rho)\ket=0$. 

In order to quantify the effect of hydrodynamical noise, 
one must investigate two-point
auto-correlations of hydrodynamical variables.
For the sake of numerical simplicity, in this work 
we focus on the two-point auto-correlation of radial flow velocity $u_r$
of the same $\tau$ (equal time) and $r$ (equal radius), which
is determined by the equal-time two-point auto-correlations of
modes $\bra v_l^s(\rho)v_l^s(\rho) \ket$.
We have checked that our conclusions are not changed from 
the analysis of other types of 
two-point auto-correlations, such as the auto-correlations of 
temperature fluctuations which depends on $\bra \delta_l(\rho)\delta_l(\rho)\ket$. 
We thereby define the following correlation function that describes
the
two-point auto-correlation of radial flow velocity of the medium,
\begin{align}
\label{equ:twop-Curur}
C_{u_r u_r}(\tau, \Delta\phi, r, \phi)=&
\bra u_r(\tau, r, \phi) u_r(\tau, r, \phi+\Delta \phi)\ket
-\bra u_{rb}(\tau, r)^2\ket\,.
\end{align}
Note that trivial contributions from the azimuthally symmetric background flow are 
subtracted in \Eq{equ:twop-Curur}. One can further focus on the correlation 
structure with respect to the relative azimuthal angle $\Delta \phi$ by 
integrating over $r$ and $\phi$,
\begin{align}
\label{equ:twop-Cs}
C_{u_r u_r}(\tau, \Delta\phi)=&
\int \frac{\tau rdr d\phi}{2\pi}
\bra u_r(\tau, r, \phi) u_r(\tau, r, \phi+\Delta \phi)\ket
-\int \frac{\tau rdr d\phi}{2\pi}\bra u_{rb}(\tau, r)^2\ket\nonumber\\
=&C_{u_ru_r}^T(\tau, \Delta\phi) + C_{u_ru_r}^I(\tau, \Delta\phi)\,.
\end{align}
The angular structure in \Eq{equ:twop-Cs} 
depends not only on the hydrodynamical noise, but also contains a fraction 
induced from initial state fluctuations, which we denote as 
$C_{u_ru_r}^T$ and $C_{u_ru_r}^I$ respectively.
The significance of hydrodynamical noise
in heavy-ion collisions is then captured by the relative contributions from 
$C_{u_ru_r}^T$ and $C_{u_ru_r}^I$. Therefore, in our numerical results, 
%to have quantitative estimate
%of the effec of hydrodynamical noise, 
we shall concentrate on the ratio between these two contributions, 
$C^T/C^I$.
%which we refer to
%as the rescaled radial flow velocity correlation funtion,
%\be
%\label{equ:res_C}
%\mbox{Recaled } C^T_{u_ru_r}(\tau, \Delta \phi) = C_{u_ru_r}^T(\tau,\Delta \phi)
%/ \mbox{magnitude of } C^I_{u_ru_r}\,.
%\ee

%We carry out numerical simulations with
%respect to these colliding systems, with a finite mount of events. For each evet,
%hydrodynamical noise is introduced by the solving Langevin equation \Eq{equ:bjo-fluc-eom} 
%for each mode
%according to the fluctuation-dissipation relation. 

%Out of phenomenological interests, 
It should be emphasized that the separation of 
two-point radial flow velocity auto-correlation into one term 
from initial state fluctuations 
and the other induced by hydrodynamical noise in \Eq{equ:twop-Cs} 
is one particular feature in our analysis, as has been demonstrated 
in the formal solution \Eq{eq:2point-k}.
In practical analysis, one is thus allowed to simulate the system evolution
independently in cases with only initial state fluctuations,
and in cases with only hydrodynamical fluctuations. 
However, in more involved studies where contributions to two-point
correlations from initial state fluctuations and hydrodynamical fluctuations
are not simply separable, one must carry out simulations with both
sources of fluctuations considered simultaneously. 
In this work, we follow the conventional procedure of solving noisy
hydro equations of motion, %with respect to initial conditions.  
by initializing the system with initial state fluctuations in
two extreme scenarios: One with a specified initial azimuthal anisotropy
and the other with a Dirac delta function. By doing so, we claim that
effects of hydrodynamical noise 
can be estimated with respect to the experimentally measured 
long-range and short-range 
correlation structures respectively.

\subsubsection{Effects of hydrodynamical noise on long-range correlations} 
%and initial state anisotropy $\varepsilon_n$}

\begin{figure}
\begin{center}
\includegraphics[width=0.45\textwidth]{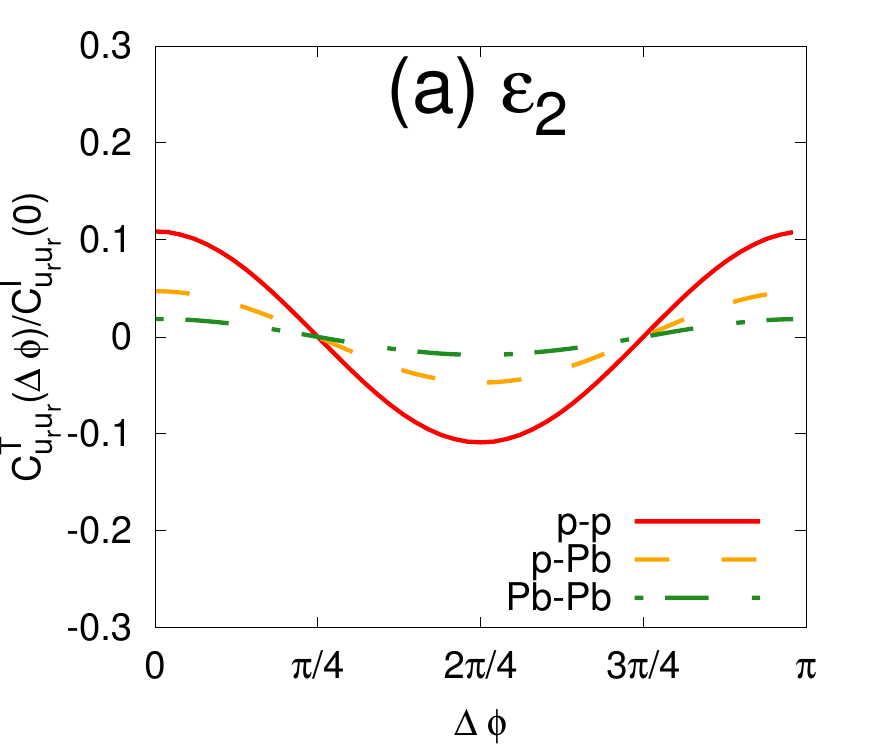}
\includegraphics[width=0.45\textwidth]{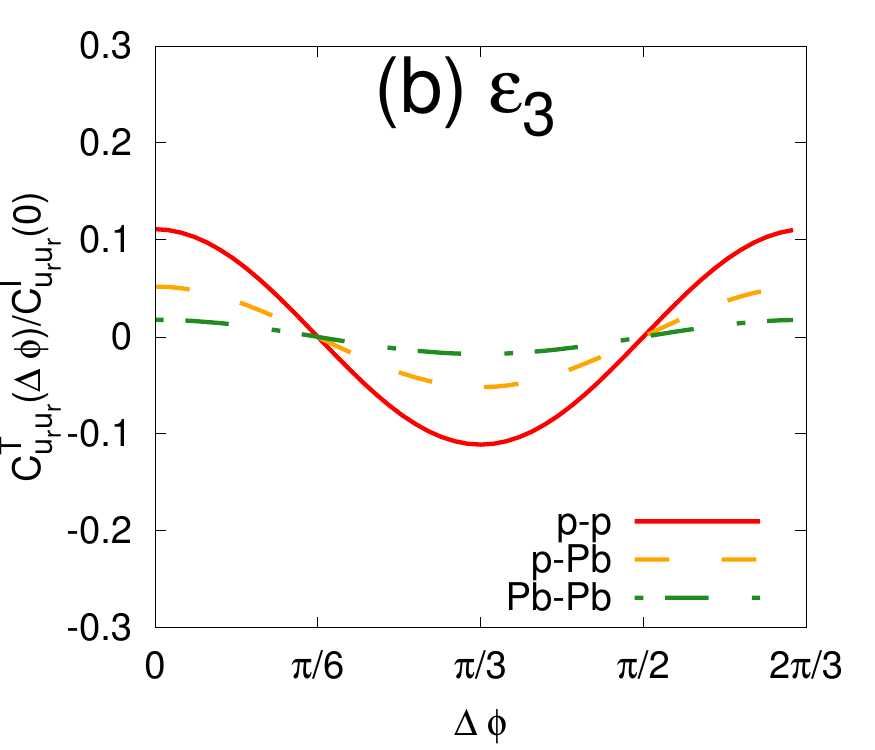}
\includegraphics[width=0.45\textwidth]{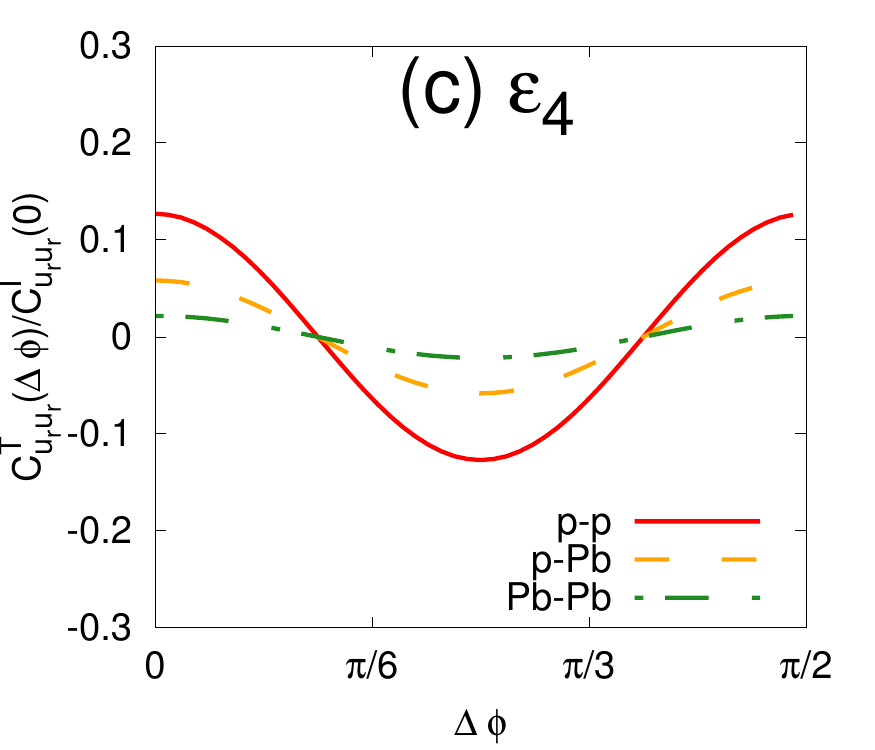}
\includegraphics[width=0.45\textwidth]{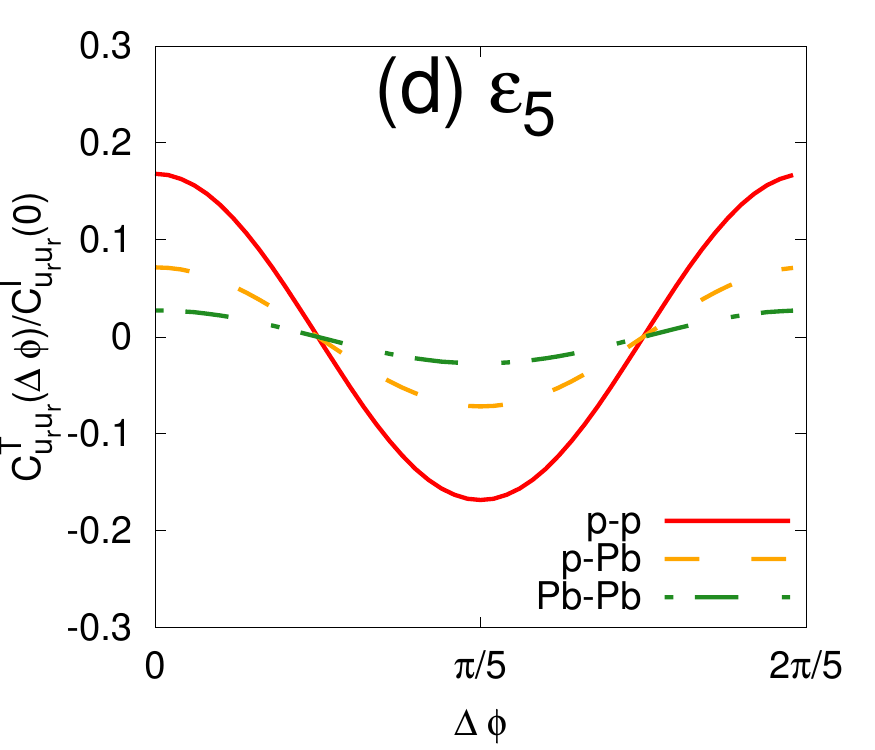}
\caption{\label{fig:2p-en}
(Color online)
Structure of the radial flow velocity auto-correlation determined by
hydrodynamical noise,
dipicted as the ratio between $C_{u_ru_r}^T(\tau, \Delta \phi)$ and 
the magnitude of $C_{u_ru_r}^I(\tau, \Delta\phi)$ 
(i.e, the absolute value of $C_{u_ru_r}^I(\tau, \Delta\phi)$ at $\Delta \phi=0$)
%which is taken be 
%the value of $C_{u_ru_r}^I(\tau, \Delta\phi)$ at $\Delta \phi=0$,
at $\tau=2.5$ fm. Results are obtained from numerical simulations of 10000 
events with respect to initial condition 
\Eq{equ:lc-ini} with an 
(a) $\varepsilon_2$, (b) $\varepsilon_3$, (c) $\varepsilon_4$ and
(d) $\varepsilon_5$, for p-p (red solid lines), p-Pb (orange dashed lines)
and Pb-Pb (green dash-dotted lines) systems.
}
\end{center}
\end{figure}

Long-range correlations of the observed spectra in heavy-ion collisions are 
associated with harmonic flow, which in turn depends on the evolution of
azimuthal anisotropies of density profile. 
For each harmonic order $m$, anisotropy is characterized by the so-called
eccentricity $\varepsilon_m$ which represents an invariant deformation
of the density profile under azimuthal rotation $\phi\rightarrow\phi+2\pi/m$.
Therefore,
we perturb the azimuthally symmetric initial profile of Gubser flow by certain %corresponding
azimuthal modes. Following the discussions originated in \cite{Gubser:2010ui},
for each harmonic order $m$, when initial state fluctuations are characterized by the
following non-zero modes
\footnote{
The $(-1)^m$ factor in the brackets originates from our convention of spherical harmonics.
}
\be
\label{equ:lc-ini}
\frac{\delta \hat T(\theta,\phi,\rho_0,\xi)}{\hat T(\rho_0)}
=
-\sqrt{\Lambda_{ini}}\left[(-1)^m\frac{1}{\sqrt{2}} Y_{m,m}(\theta,\phi) 
+ \frac{1}{\sqrt{2}} Y_{m,-m}(\theta,\phi)
\right]\,,
\ee  
one accordingly realizes a density profile with a non-zero $\varepsilon_m$.
It should be emphasized that the eccentricity generated from \Eq{equ:lc-ini} is 
closely related to the eccentricity defined in terms of cumulants with a
$r^m$ weighting~\cite{Teaney:2010vd}, yet to an approximate level. More 
detailed analysis of \Eq{equ:lc-ini} with respect to phenomenology
can be found in \cite{Gubser:2010ui}. 
The sign convention in \Eq{equ:lc-ini} is taken so that the gradient of the
deformed density is maximal along $x$-axis. 
$\Lambda_{ini}$ in \Eq{equ:lc-ini} reduces to a constant parameter to be fixed
by the values of $\varepsilon_m$ in the colliding systems.  
In our simulations with respect to the ultra-central Pb-Pb and p-Pb
colliding systems, we take $\varepsilon_2(\mbox{Pb-Pb})\sim0.05$ and 
$\varepsilon_2(\mbox{p-Pb})\sim0.15$ at $\tau=0.6$ fm, 
which are typical values considered 
in phenomenological studies of heavy-ion collisions 
(cf.~\cite{Luzum:2012wu,Bzdak:2013zma}). For ultra-central p-p collisions,
due to the smaller multiplicity production we take a larger value
of initial anisotropy $\varepsilon_2(\mbox{p-p})\sim0.2$.
We have checked that the same values of $\Lambda_{ini}$ result 
in eccentricities of higher harmonics (up to $\varepsilon_5$) of the similar
order of magnitude in the corresponding systems.

For each of the particular deformations introduced in the initial state,
we solve for the noisy Gubser flow and calculate the correlation function 
defined in \Eq{equ:twop-Cs}. Since we are only 
interested in the effect of hydrodynamical
noise on the long-range correlations, i.e., the evolution of azimuthal 
anisotropies, in the mode summation we ignore contributions from modes 
other than those of relevance to the corresponding initial anisotropies.
For instance, if one calculates the evolution of ellipticity, to
quantify \Eq{equ:twop-Cs} the mode summation only involves
$\bra (v_{2,2}^s(\rho))^2\ket$ and $\bra (v_{2,-2}^s(\rho))^2\ket$.
%Although in principle all the modes are affected by hydrodynamical noise
%and need to be taken into account in the mode summation, 
%anisotropies receives extra constrains from rotation symmetry.
%Besides higher modes in the mode summation are more influential 
%to the short-range structure, which we shall discuss later. 

In \Fig{fig:2p-en}, two-point auto-correlation 
$C_{u_ru_r}^T(\tau,\Delta\phi)$ are plotted as a function 
of $\Delta \phi$ at $\tau=2.5$ fm.
Although the structure of $C^I_{u_ru_r}(\Delta \phi)$
is not shown, it is worth mentioning that $C^I_{u_ru_r}$ and $C^T_{u_ru_r}$ share
the same
structures as a function of $\Delta \phi$, but are different in magnitudes.
The periodic correlation structures shown in \Fig{fig:2p-en} are rooted 
in the azimuthal symmetries considered for each of these cases.
%in the azimuthal anisotropies introduced in the initial state. 
For an
initial $\varepsilon_2$, $\varepsilon_3$, $\varepsilon_4$ and $\varepsilon_5$, 
correlations of radial flow velocity exhibit 
periodicity in $\pi$, $2\pi/3$, $\pi/2$ and $2\pi/5$ respectively. 
We take the ratio between %the magnitudes of correlation 
$C_{u_ru_r}^T$ and the magnitude of the correlation function $C_{u_ru_r}^I$,
i.e., $C_{u_ru_r}^T(\tau,\Delta \phi)/C_{u_ru_r}^I(\tau,0)$, 
so that one is allowed to read off directly the relative change of anisotropies due to
hydrodynamical noise in \Fig{fig:2p-en}. 
As can be seen in \Fig{fig:2p-en}, hydrodynamical noise
results in extra contributions to the development of azimuthal 
anisotropies, which are getting stronger from 
the ultra-central Pb-Pb collision systems, to ultra-central p-Pb and p-p,
and also from lower order harmonics to higher order harmonics.
The increasing contribution from hydrodynamical noise according to
harmonic order %$n$ 
can be understood as follows:
On one hand, hydrodynamcial noise is insensitive to the harmonic order, which
is subject to the independence of index $m$ in the evolution equation (cf.
\Eqs{equ:gammas}). On the other hand, however, evolution of anisotropies of 
higher order harmonics suffers stronger viscous suppression. Accounting for
both effects, one would expect that relatively hydrodynamical noise becomes
more important for higher order harmonics.
Nonetheless, the over-all magnitude of enhancement is not significant,
in particular for the Pb-Pb systems in which it is less than 3\%.

\subsubsection{Effects of hydrodynamical noise on short-range correlations}

For the analysis of short-range correlations, initial condition is chosen with temperature fluctuating 
in the transverse plane with respect to a Dirac delta function, 
\be
\label{equ:ic_delta}
\frac{\delta\hat T(\rho_0,\theta,\phi,\xi)}{\hat T(\rho_0)} = 
%\sqrt{\frac{{\tau_0^2 A_\perp}}{{9N_s}}}
\sqrt{\Lambda_{ini}}\times 
\frac{1}{\cosh\rho_0\sin\theta}\delta(\theta-\theta_0)\delta(\phi-\phi_0)\,.
\ee
%which determines $\delta_{l}(\rho_0)$. 
One may check that on an event-by-event basis, 
\Eq{equ:ic_delta} is consistent with a Dirac delta in the transverse
plane in the original Milne space-time.\footnote{
To see so, one must integrate over $\theta_0$ and $\phi_0$, which
is equivalent to average over collision events.} % an event-by-event basis.}
The same values of $\Lambda_{ini}$ fixed by eccentricities in the previous section 
are kept regarding the colliding systems under consideration. 
%Number of independent sources $N_s$ scales approximately with number of participants,
%for which we consider $N_s=N_p/2$~\cite{}.
%We take $v^s_l(\rho_0)=0$ since
%flow velocity fluctuations are neglected in this work in the initial state.
Without losing generality, we take $\theta_0=0.2$ and $\phi_0=\pi$
throughout this work, which corresponds to a peak near the origin on the x-axis.
Since higher $l$-modes receive stronger viscous suppression,
our simulations are limited to the modes $l<30$. 

\begin{figure}
\begin{center}
\includegraphics[width=1.\textwidth]{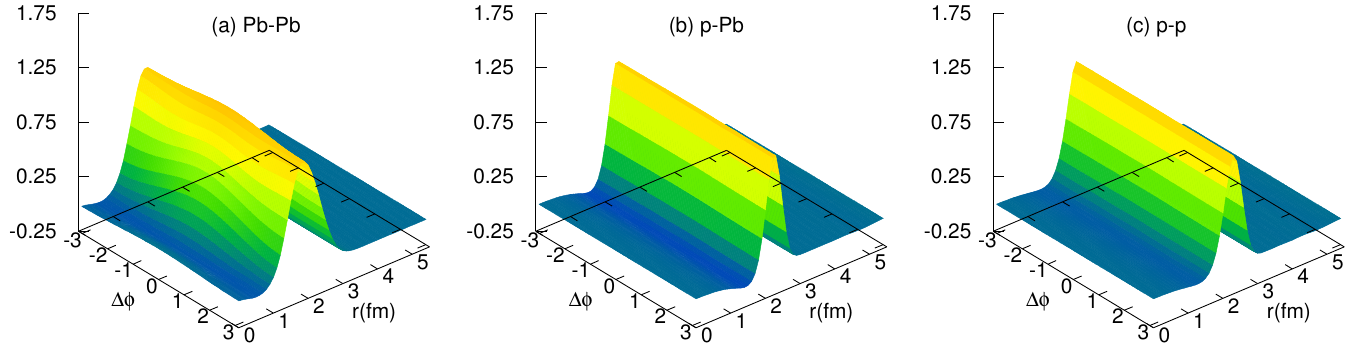}
\includegraphics[width=1.\textwidth]{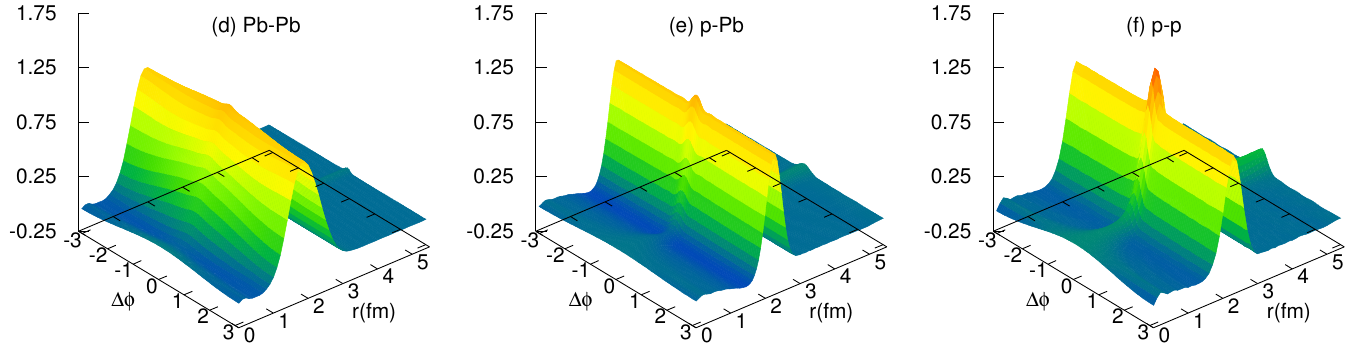}
\caption{\label{fig:2p-dU}
(Color online)
\label{fig:2d-ur2}
Two-point auto-correlations of radial flow velocity
defined by only an integration over $\phi$ in \Eq{equ:twop-Curur}, 
$C_{u_r u_r}(\tau, \Delta \phi, r)$,
%= \int \frac{d\phi}{2\pi} C_{u_r u_r}(\tau, \Delta \phi, r,\phi)$,
as a function of $r$ and $\Delta \phi$. Results are obtained 
with respect to initial condition
\Eq{equ:ic_delta} with $\theta_0=0.2$ and $\phi_0=\pi$, at $\tau=2.5$ fm,
without hydrodynamical noise (upper row) and with 
hydrodynamical noise (lower row) 
from numerical simulations of 5000 events.
To have all the three systems plotted on the same scale,
magnitudes of the correlations are rescaled in a way that
the height of the bump at $|\Delta \phi|=\pi$ is equal to 1.
}
\end{center}
\end{figure}
\Fig{fig:2p-dU}(a), (b) and (c) depict the corresponding results 
of the correlation function $C_{u_ru_r}(\tau,r,\Delta\phi)$ (without $r$ integration) in the 
ultra-central Pb-Pb, p-Pb and p-p colliding systems respectively without the
inclusion of hydrodynamical noise, at $\tau=2.5$ fm.
Regarding an initial density profile which is perturbed by a delta function,
hydro evolution results in sound-wave propagation
along with the medium expansion, which
is exactly seen as a bump in the over-all structure of $C_{u_ru_r}$.  
To have the correlation functions plotted on the same scale, we have rescaled 
the height of the bump %by a factor of $9.84$, $0.66$ and $0.51$ for the 
%Pb-Pb, p-Pb and p-p systems respectively, 
%so that the height of the bump is unity 
so that it is unity
at $|\Delta \phi|=\pi$. 
This rescaling 
%is physically subject to the different 
%expansion rate in the three systems, which 
reveals a generic feature of Gubser's solution that the medium 
expands much faster in smaller systems p-Pb and p-p
than in Pb-Pb, which however does not affect
our analysis of the effect of hydrodynamical noise.
The position of %the bump peak along $r$, namely, 
the sound horizon reflects 
hydro response to the position of the initial
delta function. %which is around $1.9$ fm for the Pb-Pb, $2.3$ fm for both p-Pb and pp. 
The shape of the bump is altered from a delta function
by diffusion and dissipation.
As one would expect, hydrodynamical noise affects the fine structure of the fluid
system, which is then reflected as an excess around $\Delta \phi=0$.
As seen in \Fig{fig:2p-dU}(d), (e) and (f), 
when hydrodynamical noise is considered in our simulations,
an excess in the auto-correlation structure indeed appears at $\Delta\phi=0$ 
and persists from around origin ($r=0$) to large
radii. Especially, on top of the bump structure 
the hydrodynamical noise leads to a near-side peak 
which is marginal in the Pb-Pb system as shown
in \Fig{fig:2p-dU}(d), but becomes significant in p-p as shown in 
\Fig{fig:2p-dU}(f).

\begin{figure}
\begin{center}
\includegraphics[width=0.5\textwidth]{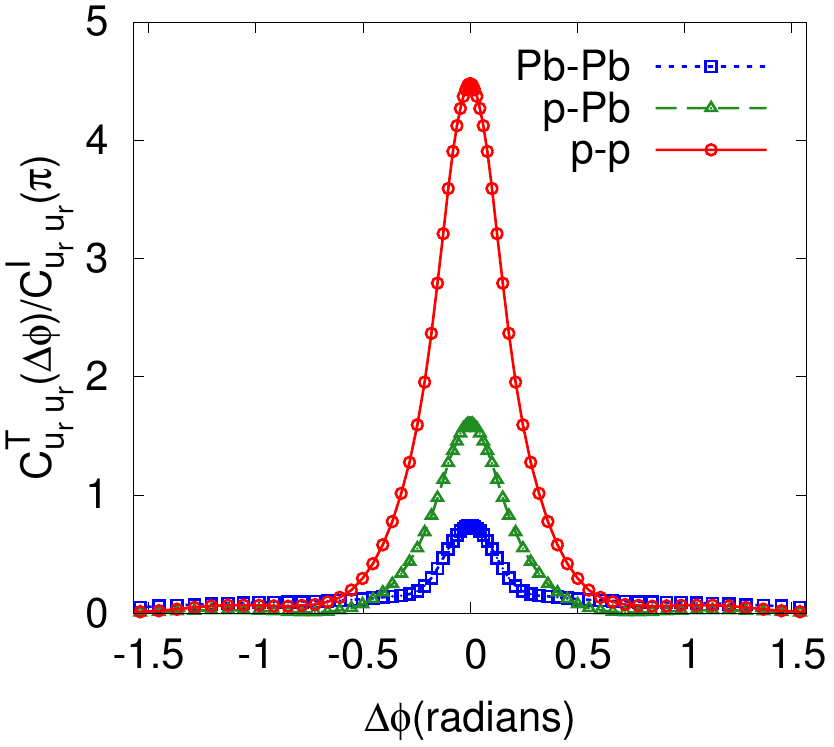}
\caption{\label{fig:2p-deltaC}
(Color online)
\label{fig:2p-rel-dU}
Details of the near-side peak in radial flow velocity correlations
due to hydrodynamical noise, depicted in terms of 
the correlation function $C_{u_ru_r}^T(\tau, \Delta \phi)$ divided by the magnitude of 
correlation due to initial fluctuation, i.e., value of $C_{u_ru_r}^I(\tau, \Delta \phi)$ 
taken at $|\Delta \phi|=\pi$, with $\tau=2.5$ fm.
}
\end{center}
\end{figure}

We next concentrate on the details of the near-side peak from the hydrodynamical
noise in \Fig{fig:2p-dU}. %In a similar manner as in \Eq{}, 
%We first integrate along $r$, but now limit the integration around the bump to
To avoid overestimation of the effect of 
hydrodynamical noise, we limit the integration over
$r$ and $\phi$ around the bump. 
%Then we obtain the correlation solely 
%from hydrodynamical noise $C_{u_ru_r}^T$ by subtracting from $C_{u_ru_r}$ the contributions 
%from the bump in \Fig{fig:2p-dU}, namely, the contributions induced from initial
%state fluctuations $C_{u_ru_r}^I$. 
Rescaled again by the magnitude of $C_{u_ru_r}^I$ (now with $\Delta\phi=\pi$), 
we obtain consequently the near-side peak in $C_{u_ru_r}^T$ due to hydrodynamical noise 
relative to the correlation structure induced by initial state fluctuations.
%\be
%\label{equ:rel-2p-dU}
%\Delta C_{u_ru_r}(\tau, \Delta \phi)=\frac{\int_{\mbox{\tiny bump}} rdr C_{u_ru_r}^T(\tau, r,\Delta\phi)}
%{\int_{\mbox{\tiny bump}} rdr C_{u_ru_r}^I(\tau,r,\Delta\phi)}\,.
%\ee
Results corresponding to 
the ultra-central Pb-Pb, p-Pb and p-p systems
are shown in \Fig{fig:2p-rel-dU}, which %\Fig{fig:2p-rel-dU} 
exhibit a clear trend in the three colliding systems.
The height, as well as the width, of
the peak increase from Pb-Pb, to p-Pb and p-p.
%Relatively, it can be understood that at least for the flow velocity correlations,
%hydrodynamical noise generates a
%near-side peak which is of 70\% of the near-side correlation strength from initial 
%state fluctuations in the Pb-Pb system. For a p-Pb system and a p-p system, the 
%relative size of the near-side peak from noise is enhanced to as 1.6 times 
%and 4.5 times large
%as the background bump.

\section{Summary and conclusions}
\label{sec:summary}

In this work we solved the noisy Gubser flow, by implementing hydrodynamical noise
on top of Gubser's solution to Navier-Stokes hydrodynamics.
Hydrodynamical noise
is formulated in a standard way according to the fluctuation-dissipation relations,
from which we noticed that
the absolute amplitude of hydrodynamical noise in ultra-central 
heavy-ion collisions is 
essentially determined by the total multiplicity of the collision event, 
instead of the system size as one might have anticipated.
Regarding the ultra-central
Pb-Pb, p-Pb and p-p collisions carried out at the LHC energies, we quantitatively
analyzed the 
effects of hydrodynamical noise with emphasis on long-range (evolution of 
azimuthal anisotropies) and short-range correlations. A clear trend of 
enhancement of the hydrodynamical noise was confirmed in both cases from the 
Pb-Pb system, to p-Pb and p-p systems, which as we have claimed, is mostly
due to the decrease of multiplicity, but not system size. 

Azimuthal anisotropies receive extra contributions from the hydrodynamical 
noise during medium evolution, 
which implies an increase of harmonic flow $v_n$.
In addition, higher order harmonics are found 
more sensitive to the hydrodynamical noise, which confirms the results from
more sophisticated hydrodynamical simulations~\cite{MuraseQM15}. 
However, at least from our simulations for the conformal and azimuthally
symmetric systems, the increase of anisotropies due to hydrodynamical noise
is not significant. Especially, we expect that for the ultra-central
Pb-Pb collision systems, the effect of hydrodynamical noise on 
harmonic flow $v_n$ is negligibly small. 
On the contrary, short-range structure of the fluid system is more 
affected by the effect of hydrodynamical noise, which is demonstrated as 
the appearance of a near-side peak.
It is 
understandable in the sense that higher order hydro modes, which correspond to
the finer structure of a fluid system, are dominated by contributions from
hydrodynamical noise, rather than hydro response to initial state fluctuations.
 We investigated in details the structure of 
of the peak, relative to the short-range two-point flow velocity correlations
induced by initial state fluctuations. We noticed that both the height
and the width are enhanced when the effect of hydrodynamical noise becomes larger. 

Bearing in mind that our analysis of the noisy Gubser flow is less realistic 
in several aspects, 
especially the caveat that we have no freeze-out process and there 
exists a gap between the two-point flow velocity correlation and the experimentally
measured two-particle correlations, 
it is implied from our results that simulations of viscous hydrodynamics 
without hydrodynamical noise are reliable for the investigation of harmonic
flow $v_n$, even in the smaller colliding systems like p-Pb.
However, hydrodynamical noise must be taken into account if one studies the
fine structure of the near-side two-particle correlations, which also contains informations
related to the dissipative properties of the QCD medium.
%First, hydrodynamical noise affects more on the short-range correlations. It is 
%understandable in the sense that higher order hydro modes, which correspond to
%the finer structure of a fluid system, are dominated by contributions from
%hydrodynamical noise, rather than hydro response to initial state fluctuations. 
%
%

\acknowledgements

We are grateful for many helpful discussions with Jean-Yves Ollitrault
at different stages of this work. LY also thanks Derek Teaney and Clint Young 
for useful conversations.
LY is funded by the European Research Council under
the Advanced Investigator Grant ERC-AD-267258.

\appendix

\section{Numerics}
There exist several schemes for solving a stochastic differential 
equation numerically (cf. \cite{Kloeden92}), depending on appropriate interpretations of the 
stochastic integral with respect to noise. Rewriting \Eqs{equ:gub-fluc-eom} as, 
\be
d\t\V(\rho) = -\t\Gamma(l,\rho)\t\V(\rho)d\rho + \t \K d\rho\,,
\ee 
we notice that $\t\K d\rho=dW_\rho$ represents an increment of Wiener process
$W_\rho$ which satisfies
\be
W_{\rho_0} = 0,\quad
\bra W_\rho\ket=0,\quad
\bra (W_{\rho+\Delta\rho}-W_\rho)^2\ket=|\Delta \rho| \times \mbox{factor}\,.
\ee 
The `factor' in the above relations indicates that the 
correlation strength of the Wiener process is additionally determined by the 
amplitude of two-point autocorrelation of hydrodynamical noise in \Eq{equ:mode-twop}.
Then for any time increase $\Delta \rho$, one needs to evaluate the 
stochastic integral numerically,
through the limit of Riemann sum,
\be
\Delta W = \int_\rho^{\rho+\Delta \rho} \t \K(\rho') d\rho'
=\lim_{n\rightarrow\infty}\sum_{i}^n \t \K(\bar\rho_i) \frac{\Delta\rho}{n}\,.
\ee
However, ambiguity arises when one takes $\bar\rho_i=\rho+i \Delta\rho/n$ or 
$\bar\rho_i=\rho+(i+1/2) \Delta\rho/n$ in the sum, which corresponds to the It\^o integral
and the Stratonovich integral respectively. For multiplicative noise (when 
$\t\K$ depends on $\t\V$) these two integrals lead to different values, while for 
noise that is not multiplicative (when $\t\K$ is 
independent of $\t\V$) It\^o and Stratonovich integrals 
coincide. Since hydrodynamical noise in our work is not multiplicative, 
recipes derived from either the 
It\^o integral or the Stratonovich integral can be applied equivalently. 
Indeed, we have checked that the Euler-Maruyama method and 
the Heun's method discussed in Ref.\cite{Young:2013fka} 
result in compatible solutions.

\section{Estimation of statistical error from ensemble average}

The total number of events for the ensemble average is taken 
accounting for the convergence of ensemble average, as well as 
statistical errors of our numerical simulations.
One can estimate statistical errors of ensemble averaged quantities 
from variance, which for the two-point correlations
requires the knowledge of four-point correlations.  
For the hydrodynamical noise considered in the linear order, 
there are no generic four-point correlations despite
the one from two-point auto-correlations.
%\footnote{
%Discussions on n-point correlations of hydrodynamical noise in the 
%theory beyond linear order can be found in \cite{Kovtun:2014hpa}. 
%%, and with respect to higher order
%%viscous hydrodynamics, are more involved and requires 
%}. 
Therefore, the variance of
any two-point correlation originated from the hydrodynamical noise 
is of the same order the two-point correlation itself. For instance,
for the temperature modes which are completely induced by hydrodynamical
noise, from Wick's theorem,
\be
\bra (\delta_l)^4\ket=3\bra (\delta_l)^2\ket^2\,,
\ee
one finds the variance 
\be
\mbox{Var}[\bra (\delta_l)^2\ket]=\sqrt{2}\bra (\delta_l)^2\ket\,,
\ee
and statistical error of $\bra (\delta_l)^2\ket$ is 
\be
\Delta\bra (\delta_l)^2\ket = \frac{\sqrt{2}\bra 
(\delta_l)^2\ket}{\sqrt{N_{\mbox{\small ensemble}}}}\,.
\ee
In this work, we take $N_{\mbox{\small ensemble}}=10000$ and 5000 for the 
investigation of long-range and short-range correlations respectively,
such that relatively 
the statistical error of two-point correlations of hydrodynamical noise
from our numerical 
simulations is less than $3\%$. 

\bibliography{refsbib}

\end{document}